\documentclass[12pt]{article}
\usepackage[utf8]{inputenc}
\usepackage{cite}
\usepackage{mathptmx}
\usepackage{setspace}

\usepackage{changepage}
\usepackage{dcolumn}
\usepackage[table]{xcolor}
\usepackage{pgfplots}
\usepackage{floatrow}
\usepackage{floatrow}   
\usepackage{subcaption}
\usepackage{graphicx,xcolor} 

\usepackage[margin=1in]{geometry}
\usepackage{amsmath}
\usepackage{amssymb}
\usepackage{mathptmx}
\usepackage{physics}
\usepackage{pgf}

\definecolor{mynavy}{HTML}{000080}
\definecolor{darkred}{HTML}{8B0000}
\definecolor{mygreen}{HTML}{006400}
\definecolor{mygold}{HTML}{B8860B}

\makeatletter
\newcommand\footnoteref[1]{\protected@xdef\@thefnmark{\ref{#1}}\@footnotemark}
\makeatother

\newcolumntype{d}[1]{D..{#1}}

\title{\vspace{0.5cm}\small{University “Ss. Cyril and Methodius” in Skopje \\
Faculty of Computer Science and Engineering\\ PhD Studies in Computer Science} \vspace{1cm}\\  \Huge{Cooperation dynamics of generalized reciprocity on complex networks} \vspace{0.5cm}\\ \small{PhD Disertation}\vspace{3cm}}
\author{Candidate: Viktor Stojkoski\footnote{Advisor: Acad. Ljupco Kocarev, PhD, }, MSc}
\date{Skopje, October 2020}

\begin{document}

\maketitle
\thispagestyle{empty}
\newpage
\begin{abstract}
Recent studies suggest that the emergence of cooperative behavior can be explained by generalized reciprocity, a behavioral mechanism based on the principle of “help anyone if helped by someone”. In complex systems, the cooperative dynamics is largely determined by the network structure which dictates the interactions among neighboring individuals. Despite an abundance of studies, the role of the network structure in in promoting cooperation through generalized reciprocity remains an under-explored phenomenon. In this doctoral thesis, we utilize basic tools from the dynamical systems theory, and develop a unifying framework for investigating the cooperation dynamics of generalized reciprocity on complex networks. We use this framework to present a theoretical overview on the role of generalized reciprocity in promoting cooperation in three distinct interaction structures: i) social dilemmas, ii) multidimensional networks, and iii) fluctuating environments. The results suggest that cooperation through generalized reciprocity always emerges as the unique attractor in which the overall level of cooperation is maximized, while simultaneously exploitation of the participating individuals is prevented. The effect of the network structure is captured by a local centrality measure which uniquely quantifies the propensity of the network structure to cooperation, by dictating the degree of cooperation displayed both at microscopic and macroscopic level. As a consequence, the implementation of our results may go beyond explaining the evolution of cooperation. In particular, they can be directly applied in domains that deal with the development of artificial systems able to adequately mimic reality, such as reinforcement learning.
\end{abstract}
\thispagestyle{empty}
\newpage
The doctoral thesis summarizes the findings described in the papers:

\begin{itemize}
    \item Cooperation dynamics of generalized reciprocity in state-based social dilemmas \\ \textbf{Viktor Stojkoski}, Zoran Utkovski, Lasko Basnarkov, and Ljupco Kocarev \\ Physical Review E E 97, 052305 (2018)
    \item The role of multiplex network structure in cooperation through generalized reciprocity \\
\textbf{Viktor Stojkoski}, Zoran Utkovski, Elisabeth Andr{\'e}, and Ljupco Kocarev \\
Physica A: Statistical Mechanics and its Applications 531, 121805 (2018)
\item Cooperation dynamics in networked geometric Brownian motion \\
\textbf{Viktor Stojkoski}, Zoran Utkovski, Lasko Basnarkov, and Ljupco Kocarev \\
Physical Review E 99, 062312 (2019)
\end{itemize}

\thispagestyle{empty}

\newpage
\tiny{.}
\vspace{9cm}

\large{\textit{To every unconditional cooperator, who dedicates his own time for the world to be a better place.}}
\thispagestyle{empty}
\newpage
\tableofcontents

\newpage
\section{Introduction}

The problem of cooperation arises in situations where individual decisions are at odds with the performance of the collective~\cite{Wang-2014,santos2008social}. Ever since the publication of Darwin’s epochal work~\cite{darwin1888descent}, the appearing paradox of cooperation in social dilemmas has been at the focus of the research community. The insight that all major transitions in biological evolution, from simple to complex structures, are characterized by some degree of cooperation and sacrifice, has subsequently led to major advances in the field~\cite{Smith-1973}. However, despite decades of investigation the cooperation paradigm is still regarded as one of the most challenging issues currently faced by scientists~\cite{Pennisi-2005}.

A particular setting whose study is of great value to the complex systems community are situations where an individual has repeated encounters within not necessarily the same group or interaction structures. In this context, the concepts of direct reciprocity (``help someone who has helped you before'')~\cite{Trivers-1971} and indirect reciprocity (``help someone who is helpful'')~\cite{nowak2005evolution} have been able to provide solutions for the emergence of cooperation in essentially disparate types of social dilemmas. Although initially structured for encounters resembling a prisoner's dilemma~\cite{Axelrod-1981}, both mechanisms have been extended to account for a disperse class of interaction structures that are ubiquitous in natural systems (see Refs.~\cite{van2012emergence,hauert2006synergy,killingback2002continuous,yoeli2013powering,hilbe2014cooperation,hilbe2015evolutionary,pan2015zero}).

While being of significant theoretical value, the extent to which direct and indirect reciprocity are able to explain cooperation in real-life systems has recently been put into question~\cite{vanDoorn-2012,Taborsky-2016}. The reason for this is that the application of the rules is costly (in terms of memory and processing requirements), in the sense that they demand high cognitive abilities such as recognition of the group with which an individual is engaged in reciprocal mechanisms or knowledge about the interaction outcomes. This, for example, limits the emergence of cooperation in systems where there is randomness in the interactions and the individuals do not posses the cognitive prowess to acknowledge with whom they play~\cite{Taborsky-2016}.

While being of significant theoretical value, the extent to which direct and indirect reciprocity are able to explain cooperation in real-life systems has recently been put into question~\cite{vanDoorn-2012,Taborsky-2016}. The reason for this is that the application of the rules is costly (in terms of memory and processing requirements), in the sense that they demand high cognitive abilities such as recognition of the group with which an individual is engaged in reciprocal mechanisms or knowledge about the interaction outcomes. This, for example, limits the emergence of cooperation in systems where there is randomness in the interactions and the individuals do not posses the cognitive prowess to acknowledge with whom they play~\cite{Taborsky-2016}.

To tackle this problem, the concept of \textit{generalized reciprocity}, formally defined as the rule of ``help anyone if helped by someone'', has been developed. This concept, whose roots lie within ``upstream reciprocity''~\cite{Boyd-1989,Nowak-2005} has been fully explored in Refs.~\cite{vanDoorn-2012,Taborsky-2016}. The intrinsic feature that may favor generalized reciprocity over others is that the proximate mechanism behind it may be explained by the changes of an individuals' physiological condition~\cite{Rutte-2007,Bartlett-2006,isen1987positive}. In other words, the decision of an individual whether to cooperate or not, is based on an \textit{internal cooperative state} which captures its past experience.

\begin{figure}[t]
\includegraphics[width=8.7cm]{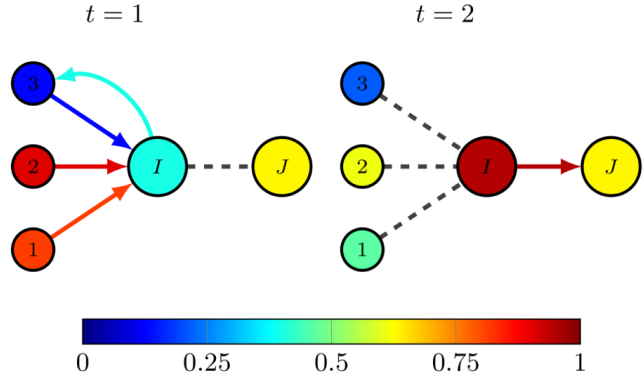}
\caption{The concept of generalized reciprocity explained by the (random) interactions within a group of 5 individuals. Solid directed lines are active relationships, whereas dashed are inactive in the respective round. In each round the individuals are colored according to their willingness for cooperation described by the internal state. As shown, in round $t=1$ individual $I$ exchanges cooperative experiences with $1$, $2$ and $3$. This results in changes in the internal states of the interacting individuals and the level of cooperation that is provided when tasked in $t = 2$. }
\label{fig:gen-rep-explanation}
\end{figure}

This simple behavioral mechanism may apply to a wide range of dynamical interaction structures. Specifically, the internal state may mimic aggregated fitness in biological systems, wealth or well-being in animal and human societies, or energy level in artificial (e.g. communication) systems. These observations are complemented with convincing empirical evidence suggesting the possibility that such a mechanism may have indeed emerged in natural systems by evolution~\cite{Rutte-2007,Leimgruber-2014,Gfrerer-2017,Bartlett-2006,Stanca-2009}.

Nevertheless, the development of theoretical models has been lacking and so far only pairwise linear interactions between individuals have been considered. Other ubiquitous scenarios that describe a social dilemma as group interactions, the possibility of nonlinear payoffs or even intertwined combinations of them, have been eschewed.

This doctoral thesis investigates the role of generalized reciprocity by performing a detailed overview of its cooperation dynamics in three different structures of interactions. The first structure is described by the notion of social dilemma which simply depicts situations where selfish behavior leads to individual gains, unless the whole population follows the same behavior~\cite{stojkoski2018cooperation}. The second structure, complements the previous study and expands the model of generalized reciprocity to multi-dimensional interactions~\cite{stojkoski2019role,stojkoski2018multiplex}. These interactions allow us to examine the presence of heterogeneous characteristics in the individual interactions, an omnipresent feature in nature. The last interaction structure evaluates the interplay between generalized reciprocity and fluctuating environments~\cite{stojkoski2019cooperation}. These environments are characterized with multiplicative growth of the accumulated payoff of an individual, and as such, have a non-trivial effect on the cooperation dynamics.

By performing a detailed analytical and numerical examination of these structures on a variety of complex network structures, among which: the Random Regular graph, the Erdos-Renyi graph, the Watts-Strogatz small world network and the Barabasi-Albert scale-free network, we show that cooperation through generalized reciprocity always emerges as the unique attractor that maximizes the global level of cooperation while at the same time preventing each participating individual from being exploited by the environment. It turns out that here a critical role plays a specific network centrality measure which quantifies the propensity of the network structure to cooperation, by dictating the degree of cooperation displayed both at microscopic and macroscopic level.

The doctoral thesis is structured as follows. In Chapter~\ref{sec:one} we give an overview of the cooperation dynamics of generalized reciprocity in social dilemmas. In this chapter, we provide a definition for a social dilemma in networked structures where the decision mechanism is based on an internal state. The definition is used to examine the thresholds which determine the dynamics of generalized reciprocity in various networked interactions. In Chapter ~\ref{sec:two}, the social dilemma definition is expanded in order to capture interactions that are happening on networks with multiple dimensions, i.e., multiplex networks. In Chapter~\ref{sec:three}, we study how the presence of a fluctuating environment affects the ability of the network structure to promote cooperation. In the last Chapter, we summarize the findings which are a result of this doctoral thesis, and provide directions for their empirical implementation.

\newpage
\section{Cooperation dynamics of generalized reciprocity in \\ social dilemmas\label{sec:one}}

\subsection{Introduction}

Motivated by the early exploits of Motro~\cite{motro1991co} on public goods, and the insights provided in~\cite{kerr2004altruism,hilbe2014cooperation,hilbe2015evolutionary}, here we revisit the concept of social dilemmas within a society that follows a state-based behavioral mechanism in the spirit of~\cite{Utkovski-2017}. As social dilemmas we consider situations in which one individual gains from selfishness, unless the whole population also behaves selfishly. Then, everyone loses. The analysis elucidates the role of the interaction topology, here described via a \textit{complex network}, which covers the standardly studied ``mixed population'' and ``regular lattice'' models as special cases. It turns out that here a critical role plays a specific \textit{network centrality measure} which quantifies the propensity of the network structure to cooperation, by dictating the degree of cooperation displayed both at microscopic and macroscopic level.

\subsection{Interaction Structure}

Formally, we consider a class of dynamical models constituted of a finite population $\mathcal{N}$ of $N$ individuals. The models run in discrete time and explain the evolution of the $N$ dimensional vector $\mathbf{p}(t) \in \left[ 0,1 \right]^N$, where the $i$-th entry, $\mathrm{p}_i(t)$, describes the internal state of individual $i$ in round $t$. According to our representation the individual payoff generated in each time step, is a function of the population state vector, 
\begin{align}
\mathrm{y}_{i,\mathcal{G}} \left( t \right) &= b_{i,\mathcal{G}} \left( \mathbf{p}(t) \right) - c_{i,\mathcal{G}} \left( \mathrm{p}_i(t) \right),
\label{eq:payoff-general}
\end{align}
where $b_{i,\mathcal{G}}$ and $c_{i,\mathcal{G}}$ are respectively the benefit and cost function of individual $i$, both parametrized by a connected graph $\mathcal{G} \left( \mathcal{N}, \mathcal{E} \right)$. The graph is defined by a set of vertices $\mathcal{N}$, corresponding to the set of individuals, and $\mathcal{E} \subseteq \mathcal{N} \times \mathcal{N}$ is the set of edges which determines the pairwise relationships between individuals.

The generality of the model is captured by the freedom in the choice of the benefit function $b_{i,\mathcal{G}}$ and the cost function $c_{i,\mathcal{G}}$. The only constraint that we thereby make is that we restrict both functions to be sufficiently smooth and adhere to several simple assumptions that define a social dilemma. In particular, we assume that the benefit function is non-decreasing with respect to each coordinate-wise projection that is not $i$, with the note that it is strictly increasing for some projections that are defined through the network topology and the rules of interaction. This implies that $i$ gains with the increase in the willingness for cooperation of a particular group of other individuals. We name this group the $l$-neighborhood of $i$ and represent it as $\mathcal{N}^{(l)}_i$. Moreover, we restrict the cost function to be an increasing function of $\mathrm{p}_i(t)$ in order to capture the social setting of paying a higher cost for the increments in the benefits of others. Finally, both functions should satisfy $b_{i,\mathcal{G}} \left( \mathbf{0} \right) = c_{i,\mathcal{G}} \left( 0 \right) = 0$ so as indicating that nothing happens when everyone defects unconditionally.

In one possible physical interpretation, the model~(\ref{eq:payoff-general}) describes the payoff of a continuous game with deterministic (i.e fixed) interactions between the individuals at each time step and where each individual has a continuum of behavioral strategies to choose from. This is, for example, the case with the continuous iterative prisoner's dilemma. In another interpretation, (\ref{eq:payoff-general}) provides a deterministic approximation for the steady state of stochastic interactions among the individuals, with random payoffs generated by the individual internal states. In this context, $b_{i,\mathcal{G}}$ and $c_{i,\mathcal{G}}$ are affine maps with respect to the random variables. The second interpretation, that we will closely follow, is aligned to the concept of generalized reciprocity, in the sense that it describes the individual payoffs as a function of the benefit they obtain through random interactions with individuals from their neighborhood, without explicit records to the contributions during these interactions.

\subsection{The social dilemma}

The definitions of the benefit and cost functions offer a realistic representation for a plethora of real-life situations. However, they alone do not describe the social dilemma. To set the stage for this phenomena, we must define two additional conditions that have to be present within the payoff structure.

First, for every individual there should exist a point beyond which cooperation is costly, i.e. beyond this point the individual is better-off by not increasing the willingness to cooperate, while expecting cooperation from others. This implies that, full cooperation by all individuals does not belong to the set of Nash equilibria, defined as the set of points that maximize the individuals' payoff given the set of available actions of all other individuals. Formally, we define this condition as the point $\mathbf{p} \in \left[ 0,1 \right]^N $, such that for every other point $\mathbf{\hat{p}}$ satisfying $\mathrm{proj}_j \left(\mathbf{\hat{p}} \right) \geq \mathrm{proj}_j \left(\mathbf{p} \right)$ for all $j \in \mathcal{N}^{(l)}_i$, where $\mathrm{proj}_j \left(\mathbf{\hat{p}} \right)$ is the $j$-th coordinate-wise projection of $\mathbf{\hat{p}}$ , the rate at which the benefit of individual $i$ changes is lower than the change in the cost when only its internal state is slightly perturbed while everything else is kept constant. In other words, for all those $\mathbf{\hat{p}}$,
\begin{align}
\pdv{b_{i,\mathcal{G}} (\mathbf{\hat{p}})}{\mathrm{p}_i} < \dv{c_{i,\mathcal{G}}(\mathrm{\hat{p_i}})}{\mathrm{p}_i}.
\label{eq:dilemma_condition_1}
\end{align}

Second, for all $\mathbf{\hat{p}} \in \left[ 0,1 \right]^N $, the sum over the changes in benefits of all $j \in \mathcal{N}$ must be greater than or equal to the change in cost provided by $i$,
\begin{align}
\sum_j \pdv{b_{j,\mathcal{G}} (\mathbf{\hat{p}})}{\mathrm{p}_i}  \geq \dv{c_{i,\mathcal{G}} (\mathrm{\hat{p_i}})}{\mathrm{p}_i},
\label{eq:dilemma_condition_2}
\end{align}
with the strict inequality holding when $\mathbf{\hat{p}} = \mathbf{0}$.

Condition~(\ref{eq:dilemma_condition_2}) together with the definitions of the benefit and cost functions, $b_{i,\mathcal{G}}$ and $c_{i,\mathcal{G}}$, reveals that full cooperation by all individuals, $\mathrm{p}_i(t) = 1$, for all $i$ and $t$, is an efficient solution, i.e. it is a solution that maximizes the overall population payoff. Considered together these two conditions build a structure under which egoistic behavior leads to depletion of the performance of the overall system, which is exactly the social dilemma metaphor.

A typical example for an interaction mechanism that is easily captured by this representation is the snowdrift game, where cooperation is a favorable trait for an individual as long as the number of cooperative neighbors is low~\cite{Santos-2005}. Another group of mechanisms that are modeled when condition (\ref{eq:dilemma_condition_1}) holds for all $\mathbf{\hat{p}} \in \left[ 0,1 \right]^N$ and condition (\ref{eq:dilemma_condition_2}) is a strict inequality for every $\mathbf{\hat{p}}$ are the prisoner's dilemma~\cite{Santos-2005}, the common pool resource problem~\cite{ostrom2015governing}, the public goods game~\cite{Perc-2017}, as well as extensions and combinations of these three interactions structures that constrain the cost of an individual to be a function only of its own state are covered (see for example~\cite{milinski2012interaction,motro1991co,hilbe2014cooperation,hilbe2015evolutionary,Doebeli-2005,archetti2012game,battiston2017determinants}).

\subsection{Behavioral update mechanism}

In our scenario, the individuals follow the state-based behavioral update introduced in~\cite{Utkovski-2017}, apart for a fraction of the population (a set $\mathcal{D}$) of unconditional defectors (with $\mathrm{p}_i (t) = 0$, $i\in \mathcal{D}$, for all $t$). 

The behavioral update rule describes the cooperative state of individual $i$ at time $t+1$ as a function of its' accumulated payoff $\mathrm{Y}_{i,\mathcal{G}}(t)$ by time~$t$ 
\begin{align}
\mathrm{p}_i(t+1)=f_i \left(\mathrm{Y}_{i,\mathcal{G}}(t)\right),
\label{eq:update}
\end{align}
where $\mathrm{Y}_{i,\mathcal{G}}(t)=\mathrm{Y}_{i,\mathcal{G}}(t-1)+\mathrm{y}_{i,\mathcal{G}}(t)$, with $\mathrm{Y}_{i,\mathcal{G}}(0)$ being the initial condition and $\mathrm{y}_{i,\mathcal{G}}(0)=0$. 

We assume that the function $f_i :\mathbb{R} \to \left(0, 1 \right)$, which maps the payoff $\mathrm{Y}$ to the willingness for cooperation $\mathrm{p}_i$, is continuous on the interval $\left(0, 1 \right)$ and has a continuous inverse (i.e. is a homeomorphism). Additionally, it is monotonically increasing with $\lim_{\mathrm{Y}\to -\infty} f_i(\mathrm{Y}) = 0$ and $\lim_{\mathrm{Y}\to \infty} f_i(\mathrm{Y}) = 1$. 

An example of a function with the above properties (which, for example, is often used for modeling in biology and ecology) is the logistic function $f_i(\mathrm{Y})=\left[1+e^{-\kappa_i(\mathrm{Y}-\omega_i)}\right]^{-1}$, where the parameters $\kappa_i$ and $\omega_i$ define the steepness, respectively the midpoint of the function.

This rule provides a simple description for the cooperative behavior in a wide range of dynamical interaction structures. Specifically, the internal state may mimic aggregated fitness in biological systems, wealth or well-being in animal and human societies, or energy level in artificial systems~\cite{Utkovski-2017}. Its advantage lies in its simplicity since (\ref{eq:update}) can easily be described as a Markovian process where an individual only has to know its present state and payoff in order to determine the next action. This is significantly different from other reciprocal update rules. For instance, in certain interaction structures direct reciprocity requires extensive memory requirements to record own and opponents' actions in order for cooperation to thrive~\cite{hilbe2017memory}.

\subsection{Cooperation dynamics}

We begin the analysis by studying the properties of the steady state solution $\mathbf{p}^*$. We will thereby assume that the interaction structure is non-degenerate, which, in our terms, means that the Jacobian of the individual payoff functions accounting only for the individuals that follow the generalized reciprocity update rule, $\mathbf{J_y}^{\setminus\mathcal{D}}(\mathbf{p})$, at the point $\mathbf{0}$ is nonsingular. In fact, it is easy to notice that as a consequence of~(\ref{eq:dilemma_condition_2}), this assumption always holds when the whole population follows~(\ref{eq:update}).

Now, in steady state, for the individuals adhering to (\ref{eq:update}), i.e. the non-defectors, it holds 
\begin{align}
\mathrm{p}^*_i&=f_i\left(f_i^{-1}(\mathrm{p}^*_i)+b_{i,\mathcal{G}} \left( \mathbf{p}^* \right) - c_{i,\mathcal{G}} ( \mathrm{p}^*_i)\right)
\label{eq:iterative_steady_state}.
\end{align}
By applying the inverse map, we obtain
\begin{align}
f_i^{-1}\left(\mathrm{p}_i^*\right)&=f_i^{-1}\left(\mathrm{p}_i^*\right)+b_{i,\mathcal{G}} \left( \mathbf{p}^* \right) - c_{i,\mathcal{G}} \left( \mathrm{p}^*_i \right).
\label{eq:steady_state_individual_inverse}
\end{align}
For the above equation to hold, it is required that 1)~$\mathrm{y}^*_{i,\mathcal{G}}\doteq b_{i,\mathcal{G}} \left( \mathbf{p}^* \right) - c_{i,\mathcal{G}} \left( \mathrm{p}^*_i \right)=0$, unless either 2)~$\mathrm{p}_i^*=1$ (i.e. $\mathbf{\mathrm{Y}}_{i,\mathcal{G}}^*=\infty$), or 3)~$\mathrm{p}_i^*=0$ (i.e.$\mathbf{\mathrm{Y}}_{i,\mathcal{G}}^*=-\infty$). 

 It is easy to verify that condition 3) is a special case of condition 1) when $\mathrm{p}^*_j=0$ for all $j \in \mathcal{N}^{(l)}_i$. Indeed, when $\mathrm{p}^*_i=0$, it must be that $y^*_{i,\mathcal{G}}\leq 0$. By the definition of (\ref{eq:update}), the condition $y^*_{i,\mathcal{G}} < 0$ implies $\mathrm{p}^*_i=0$ which, on the other hand implies $y^*_{i,\mathcal{G}} \geq 0$, leading to a contradiction. Therefore, whenever $\mathrm{p}^*_i=0$ it must hold that $ y^*_{i,\mathcal{G}} = 0$, which is true if and only if $\mathrm{p}^*_j=0$ for all $j \in \mathcal{N}^{(l)}_i$.

These conditions, together with the (strict) monotonicity of the function $f_i$ and the assumption that the interactions are non-degenerate, reveal that each individual $i$ conforming to (\ref{eq:update}) will increase, respectively decrease, its willingness for cooperation (based on its environment, i.e. on the state dynamics of other individuals), until the system reaches a steady state where it is either satisfied $\mathrm{y}^*_{i,\mathcal{G}} = 0$ or $\mathrm{p}^*_{i} = 1$, for all $i \not\in \mathcal{D}$. 
In other words, each individual following the generalized reciprocity rule will cooperate with the maximal willingness, given its environment, so as it is not exploited. In this context, the steady state $\mathbf{p}^*$ may be interpreted as a solution of the constrained optimization problem of maximizing the global level of cooperation, subject to all individuals that follow the rule receiving a non-negative payoff while the unconditional defectors have $\mathrm{p}^*_i = 0$,
\begin{equation}
\begin{aligned}
\mathbf{p}^*  &= \underset{\mathbf{p} \ \in \ \left[0,1 \right]^N}{\text{arg max}}\left \{ \sum_{i\not\in\mathcal{D}} \mathrm{p}_i; \; \mathrm{y}_{i,\mathcal{G}}  \left( \mathbf{p}\right) \geq 0 \; \forall i \ \texttt{and} \ \mathrm{p}_i = 0 \; \forall i \in \mathcal{D} \right \}.
\end{aligned}
\label{eq:optimization-solution}
\end{equation}
In our system there exists only one solution to~(\ref{eq:optimization-solution}). This implies that local cooperative behavior diffuses as a flux of energy over the network, eventually saturating at a point where cooperation is maximized under the constraint that no single individual is exploited.

\subsection{Phase transitions}

Next, we turn our attention to the phase transitions of the system. There are two such points which are of particular interest to us. The first transition corresponds to the situation when the behavioral rule  (\ref{eq:update}) is not able to support cooperation, i.e. below this point the steady state solution is described by $\mathbf{p}^* = \mathbf{0}$. 
The contrapositive of this condition represents a weak requirement for cooperation, in the sense that it represents a necessary condition for individuals with positive willingness for cooperation to exist.

The second transition point that we investigate quantifies a stronger condition for cooperation, corresponding to the situation where above this point all individuals (aside the unconditional defectors) are unconditional cooperators, i.e. the steady-state solution reads $\mathrm{p}^*_i = 1$ for all $i \not\in\mathcal{D}$. 

We study both transitions by using a differential equation representation of the update rule 
\begin{align}
\dot{\mathrm{p}}_i =\dv{f_i \left( f_i^{-1} ( \mathrm{p}_i)\right)}{Y}  \left[ b_{i,\mathcal{G}} \left( \mathbf{p} \right) - c_{i,\mathcal{G}} \left( \mathrm{p}_i\right) \right].
\label{eq:differential}
\end{align}
In the following we provide a description of the conditions that are required for extinction and full unconditional cooperation to happen. In particular, by analyzing the asymptotic stability of the system at the fixed point $\mathbf{p}^* = 0$ we show that the sufficient condition for extinction of cooperation can be approximated as
\begin{align}
\lambda_{\max} \left( \mathbf{J^{\setminus\mathcal{D}}_y}(\mathbf{0}) \right) < 0.
\label{eq:coop_extinciton_threshold}
\end{align}
where $\lambda_{\max}(\mathbf{J^{\setminus\mathcal{D}}_y}(\mathbf{p}))$ is the largest eigenvalue of~$\mathbf{J^{\setminus\mathcal{D}}_y}(\mathbf{p})$.

In fact, a similar result can be reached by studying the optimization problem~(\ref{eq:optimization-solution}). Altogether, this indicates that when every eigenvalue of the reduced Jacobian $ \mathbf{J^{\setminus\mathcal{D}}_y}(\mathbf{0})$ is negative, the system becomes dissipative and there is a continuous shrinkage in the displayed level of cooperation.

Similarly, by examining the stability of the linearized system at the point $\mathbf{1}_{\setminus\mathcal{D}}$ (the $N$ dimensional vector with entries $1$ for all $i \not\in \mathcal{D}$ and $0$ otherwise), we get that unconditional cooperation is asymptotically stable if 
\begin{align}
\min_{i \not\in \mathcal{D}} \left( v^{\setminus\mathcal{D}}_{i,\mathcal{G}} \right) > 1,
\label{eq:full_coop_threshold}
\end{align}
where $v^{\setminus\mathcal{D}}_{i,\mathcal{G}}$ is 
\begin{align}
v^{\setminus\mathcal{D}}_{i,\mathcal{G}} = \frac{b_{i,\mathcal{G}} (\mathbf{1}_{\setminus\mathcal{D}})}{ c_{i,\mathcal{G}}(1)}.
\end{align}

We note that the strict inequality in the approximation of the condition might be relaxed, as suggested by the sufficient and necessary conditions for unconditional cooperation. Indeed, on  the one hand, from the properties of the update rule we have that $v^{\setminus\mathcal{D}}_{i,\mathcal{G}} > 1$ for all $i \not\in \mathcal{D}$ implies $\mathrm{y}^*_{i,\mathcal{G}} > 0$, and hence $\mathrm{p}_i^* = 1$ for all $i \not\in \mathcal{D}$. On the other hand, in the necessary part the weak inequality $v^{\setminus\mathcal{D}}_{i,\mathcal{G}} \geq 1$ holds, which follows by directly substituting $\mathrm{p}_i^* = 1$ in (\ref{eq:payoff-general}) and repeating the same argument for all $i \not\in \mathcal{D}$. This further implies that individuals for which the inequality is not satisfied will never cooperate unconditionally. In this sense, the quantity $v^{\setminus\mathcal{D}}_{i,\mathcal{G}}$ arises as an index that quantifies the burden of each individual when cooperating, thus ultimately determining the degree of cooperation both at microscopic and macroscopic level. 

\subsection{Example}

\subsubsection{Public goods and donations}

As a constructive example for the applicative power of our framework, we consider an interaction mechanism which couples the iterated public goods game with a simple donation activity. 

In particular, we assume that in each time step nature randomly conditions whether the state of the global system is in provision of public goods or in donation. When in public goods state, each individual $i$ acts as a factor in $d_i$ processes of production represented by its nearest neighbors $\mathcal{N}^1_i$, with $d_i =|\mathcal{N}^1_i|$ being the degree of the individual. Its input in each round $t$ is proportional to the internal cooperative state $\mathrm{p}_i(t)$, whereas its productivity is inversely related with the degree $d_i$, thus leading to equal aggregate productivity of each individual.

For concreteness, we consider a linear production function,
\begin{align*}
\mathrm{q}_j(t) &= \alpha \sum_k \frac{A_{kj}}{d_k} \mathrm{p}_k ,
\end{align*}
where $\alpha$ is a parameter that describes the efficiency among the producers and $A_{kj} \in \{0,1\}$ is the $(k,j)$-th entry of the adjacency matrix $\mathbf{A}$ of the graph. In order to follow standard practice, we always fix $A_{ii} = 1$, thus each individual has one production process represented by its own vertex. To make the provision of public good related with generalized reciprocity we assume that after production takes place, $\mathrm{q}_j(t)$ is distributed to a randomly (on uniform) chosen individual from the nearest neighbors of $j$. 

On the other hand, when the system is in a donation state, each individual $i$ chooses a nearest neighbor at random and decides whether to donate him an amount $\alpha$ with probability proportional to its internal state $\mathrm{p}_i(t)$. 

The resulting public goods interaction is a randomized version of the model introduced in~\cite{santos2008social}, which has been extensively studied from the perspective of network reciprocity (a comprehensive review is provided  in~\cite{perc2013evolutionary}). In addition, when combined with the donation activity, it resembles the famous ``carrot'' mechanism in a generalized reciprocity system~\cite{milinski2012interaction}.

\subsubsection{Properties}

In this structure, the random payoff of individual $i$ in round $t$ is defined as 
\begin{align*}
\mathrm{y}_i(t)&= \mathrm{e}(t) \mathrm{y}_{1i}(t) + (1 - \mathrm{e}(t)) \mathrm{y}_{2i}(t),
\end{align*}
where $\mathrm{e}(t)$ is a Bernoulli random variable with parameter $\varepsilon$ that describes the global state, and,
\begin{align}
\begin{split}
\mathrm{y}_{1i}(t) &= \sum_{j} \mathrm{m}_{1ji}(t) \mathrm{q}_j(t) - \mathrm{p}_i(t), \\
\mathrm{y}_{2i}(t) &= \alpha \sum_{j\neq i} \mathrm{m}_{2ji}(t) \mathrm{x}_j(t) - \mathrm{x}_i(t),
\end{split}
\label{eq:carrot-payoffs}
\end{align}
are the payoffs of the public goods game and the donation process. In equation (\ref{eq:carrot-payoffs}), $\mathrm{m}_{1ji}(t)$, $\mathrm{m}_{2ji}(t)$ and $\mathrm{x}_j(t)$ are, respectively, Bernoulli random variables with parameters $A_{ji} / d_j$, $A_{ji} / (d_j-1)$ and $\mathrm{p}_j(t)$.  

In this example, the benefit and cost functions represent affine maps with respect to the random variables. Therefore, we can approximate the analysis of the steady state with the following deterministic payoff
\begin{align}
\mathrm{y}_i(t) = \varepsilon \alpha \sum_j \frac{A_{ji}}{d_j} \sum_k \frac{A_{kj}}{d_k} \mathrm{p}_k(t) + (1- \varepsilon) \alpha \sum_{k\neq i} \frac{A_{ki}}{d_k-1} \mathrm{p}_k(t) - \mathrm{p}_i(t).
\label{eq:carrot-deterministic}
\end{align}
From (\ref{eq:carrot-deterministic}) one can easily deduce that for each particular individual $i$, conditions (\ref{eq:dilemma_condition_1}) and  (\ref{eq:dilemma_condition_2}) are satisfied whenever $\alpha \in \left( 1, \frac{d_i}{z_{i}} \frac{1}{\varepsilon}\right)$, where  $z_{i} = \sum_j A_{ji}/d_j$ is a centrality measure which quantifies the expected number of times that individual $i$ is selected to be at the receiving end of a production process or a donation. The effect of the distribution of $z_{i}$ (also referred to as neighborhood importance index) on the network propensity for cooperation has been discussed in \cite{Utkovski-2017}.  

A refined version of this index, which is proportional to the maximal average benefit of individual $i$, can be defined as 
\begin{align}
s^{\setminus\mathcal{D}}_{i} = \varepsilon \sum_j \frac{A_{ji}}{d_j} \sum_{k\not\in\mathcal{D}} \frac{A_{kj}}{d_k} + (1 - \varepsilon) \sum_{k\not\in\mathcal{D}\cup i} \frac{A_{ki}}{d_k-1}.
\label{eq:pgodds-index}
\end{align}
 Its relation with the efficiency parameter $\alpha$ directly determines the level of cooperation of the system because the cooperation index $v^{\setminus\mathcal{D}}_{i,\mathcal{G}}$ can be written as 
\begin{align*}
v^{\setminus\mathcal{D}}_{i,\mathcal{G}} &= \alpha s^{\setminus\mathcal{D}}_{i}.
\end{align*}
This implies that full cooperation will exist whenever
\begin{align}
\alpha > \frac{1}{s^{\setminus\mathcal{D}}_{\min}},
\label{eq:full-coop-pgoods}
\end{align}
where $s^{\setminus\mathcal{D}}_{\min}$ is the minimum among the $s^{\setminus\mathcal{D}}$ indices.

Finally, the matrix $\mathbf{J^{\setminus\mathcal{D}}_y}(\mathbf{0})$ whose largest eigenvalue determines whether cooperation will die out or not is 
\begin{align*}
\mathbf{J^{\setminus\mathcal{D}}_y} (\mathbf{0}) &= \alpha \left[\varepsilon \mathbf{M_1^{\setminus\mathcal{D}}} + (1 - \varepsilon)\mathbf{M_2^{\setminus\mathcal{D}}} \right] - \mathbf{I},
\end{align*}
where the $(i,j)$-th entries of $\mathbf{M_1^{\setminus\mathcal{D}}}$ and $\mathbf{M_2^{\setminus\mathcal{D}}}$ are $M^{\setminus\mathcal{D}}_{1ij} = \sum_k \frac{A_{ki}}{d_k} \frac{A_{jk}}{d_j}$ and $M^{\setminus\mathcal{D}}_{2ij} = \frac{A_{ji}}{d_j-1}$ ( $M^{\setminus\mathcal{D}}_{2ii} = 0$), and $\mathbf{I}$ is the identity matrix.

This implies that the condition for extinction of cooperation reads
\begin{align}
\alpha < \frac{1}{\lambda_{\max}\left(\varepsilon\mathbf{M_1^{\setminus\mathcal{D}}} + (1 - \varepsilon)\mathbf{M_2^{\setminus\mathcal{D}}}\right)}.
\label{eq:extinction-coop-pgoods}
\end{align}

In the special case when every individual follows the update (\ref{eq:update}), $\mathbf{M_1^{\setminus\mathcal{D}}}$ and $\mathbf{M_2^{\setminus\mathcal{D}}}$ represent left stochastic matrices, in which case  $\lambda_{\max}\left(\varepsilon\mathbf{M_1^{\setminus\mathcal{D}}} + (1 - \varepsilon)\mathbf{M_2^{\setminus\mathcal{D}}}\right)  = 1$. Since $\alpha > 1$ is prerequisite for a social dilemma, this condition will never hold in a population consisting exclusively of individuals that follow the state-based rule (i.e. without defectors).

\subsection{Discussion}
 
We test these analytical findings on three different types of random graph models, the Random Regular (RR) graph, the Erdos-Renyi (ER) random graph and the Barabasi-Albert (BA) scale-free network. 
 
\begin{figure*}[t!]
\begin{adjustwidth}{-0.2in}{0in}
\includegraphics[width=17cm]{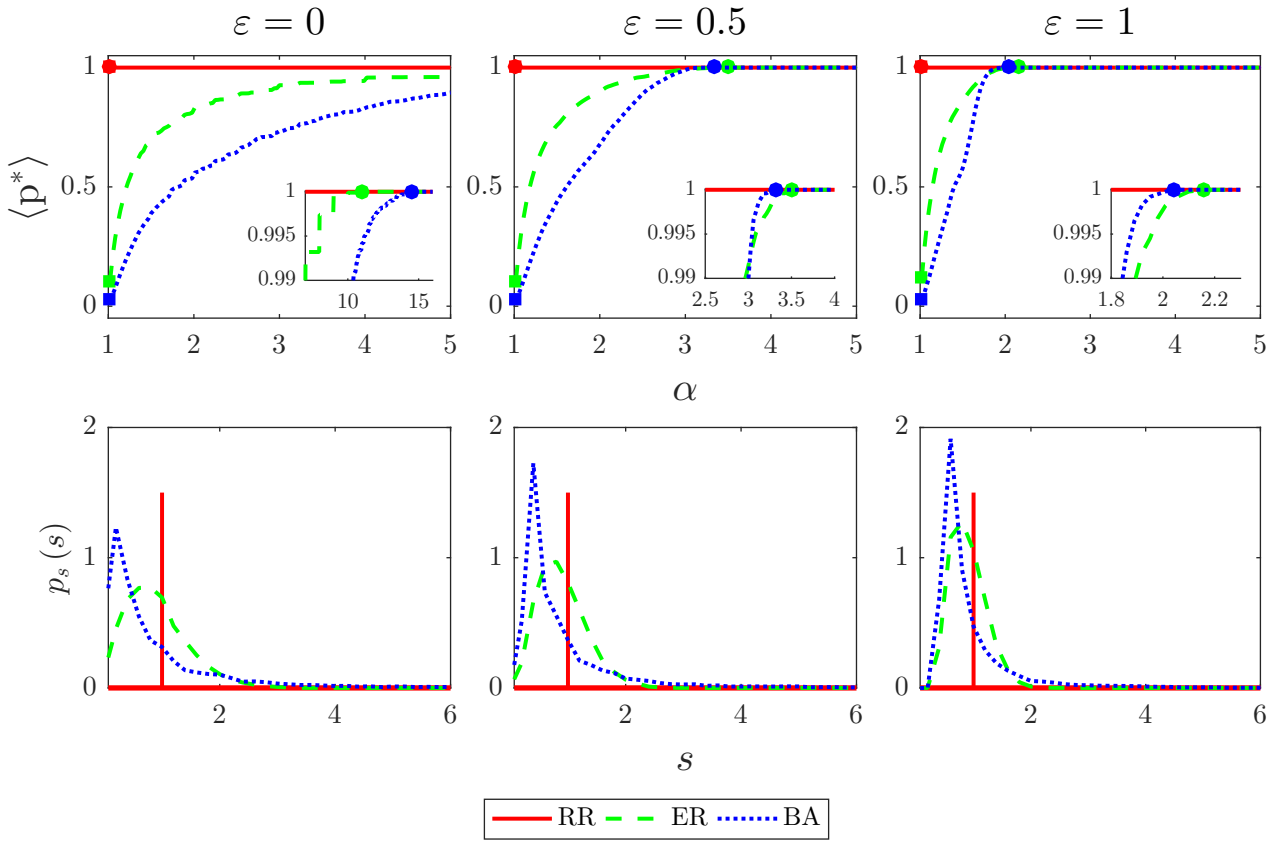}
\caption{Results for the public goods example. 
Top row shows the fraction of unconditional cooperators $\langle \mathrm{p}^* \rangle$ as a function of $\alpha$, while the bottom row gives the estimated probability density function of the index $s$. The columns correspond to different values of the parameter $\epsilon$. The results are averaged over 100 realizations with each graph having $N = 100$ and average degree $5$.
\label{fig:initial_results}}
\end{adjustwidth}
\end{figure*}

Throughout the analysis, as a measure for the global level of cooperation we consider the fraction of unconditional cooperators $\langle \mathrm{p}^* \rangle$ out of the individuals that follow the state-based update rule. In the top row of Fig.~\ref{fig:initial_results} we study the evolution of this variable as a function of the efficiency parameter $\alpha$ when $\mathcal{D} = \emptyset$ and by considering three cases of $\varepsilon$. Namely, we examine the situations when there is only a donation process~($\varepsilon = 0$), when there is equal probability for happening of both processes~($\varepsilon = 0.5$) and when only public goods are present in the system~($\varepsilon = 1$).

The condition for extinction of cooperation is plotted in the figure as a square point. 
As suggested by the analytical findings, cooperation in each graph and interaction structure exists as long as $\alpha$ satisfies the condition  for a social dilemma. 

However, the dependency of $\langle \mathrm{p}^* \rangle$ on $\alpha$ is generally disparate across the three interaction structures we explore, with the only similarity arising in the RR graph. In that case, it can be seen that full unconditional cooperation happens always independently of the interaction structure. For the ER and BA graphs, we discover that in each interaction structure the former graph acts as a better promoter of cooperation for low values of $\alpha$. While this persists when donation is the sole mechanism, whenever provisioning of public goods enters the system, a critical point appears beyond which the BA graph performs better in terms of the number of unconditional cooperators (see the inset plots).

To interpret these observations, in the bottom row of Fig.~\ref{fig:initial_results} we display the probability density function of the index $s^{\setminus\mathcal{D}}$, when there are no defectors (here simply denoted as $s$). The figure shows that the distribution of this variable in the RR graph resembles a Dirac delta function with the mass centered on $1$. This implies that the condition for asymptotic stability of unconditional cooperation is satisfied for every plausible $\alpha$, which in turn implies the full unconditional cooperation observed at $\alpha = 1$. For the ER and BA random graphs we notice that the distribution of $s$ is positively skewed with the BA graph exhibiting by far larger skewness. A direct consequence of this is the remark for the evolution of $\langle \mathrm{p}^* \rangle$ in the two graphs. More precisely, the right tail of the distribution determines the degree of cooperation provided when the level of efficiency is low. In this regard the lower slope in the ER graph indicates that for low $\alpha$, there will be more individuals for which the necessary condition for unconditional cooperation is satisfied.

Similarly, the left tail explains why the BA graph converges to full unconditional cooperation faster than the ER graph in the case of public goods. In particular, as $\varepsilon$ increases the kurtosis in the distribution of $s$ also increases for both ER and BA graphs, implying that the individuals tend to become more similar in regard to this index. The increment is larger for the BA graph, and therefore the lower thresholds for unconditional cooperation in the case of public goods.

\begin{figure*}[t!]
\begin{adjustwidth}{-0.1in}{0in}
\includegraphics[width=15cm]{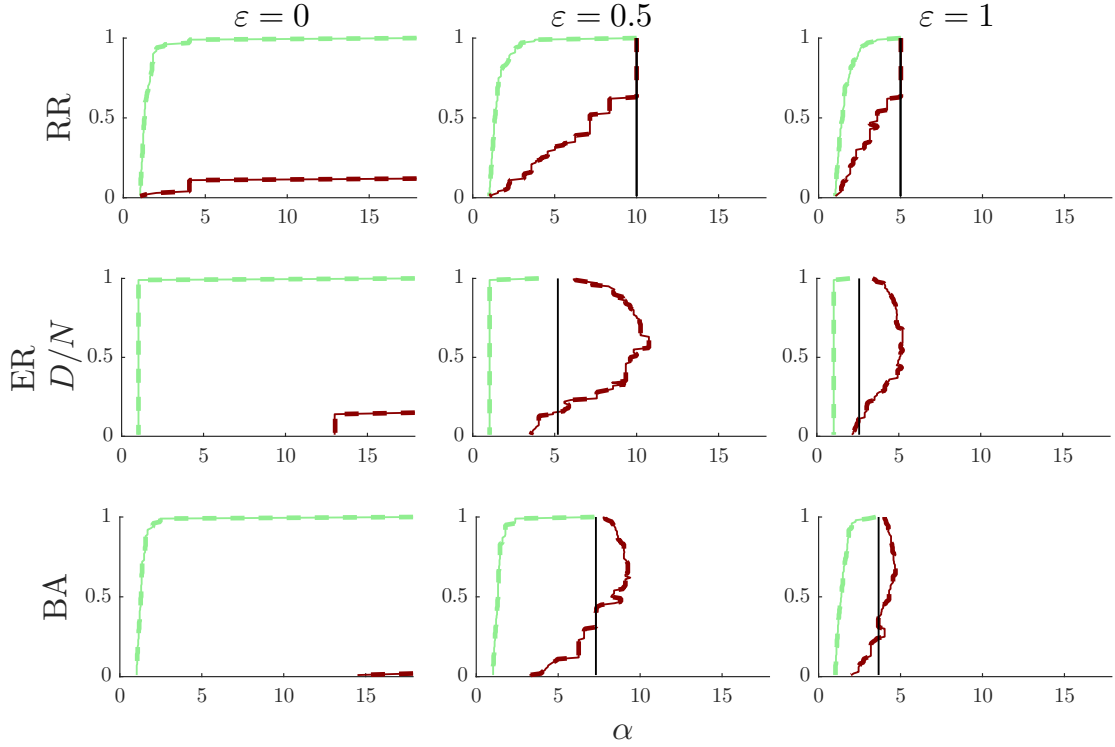}
\caption{Contour plots for the transition points. The green (light gray) curves are contour plots for the transition points to extinction of cooperation while the red (dark gray) curves are the transition points to full unconditional cooperation, with the dashed lines indicating the estimated thresholds from equations (\ref{eq:coop_extinciton_threshold}) and (\ref{eq:full_coop_threshold}). With the black vertical line we denote the minimum of $d/z$. The rows represent different types of random graphs (RR, ER and BA) while the columns correspond to different values of $\epsilon$. The results are averaged across 100 graph realizations consisting of 100 individuals with each graph having an average degree of $5$.}
\label{fig:results_defectors}
\end{adjustwidth}
\end{figure*}

In order to quantify the effect of defectors, in Fig.~\ref{fig:results_defectors} we display contour plots for the fraction of unconditional cooperators as a function of both the efficiency parameter and the fraction of individuals that are unconditional defectors for each of the studied random graphs. In the figure, above the green (light gray) lines is the region where cooperation ceases to exist, whereas below the red (dark gray) lines is the region where all individuals outside the defector set cooperate unconditionally. In the region in-between, unconditional cooperators coexist with partial cooperators (those with $0 < \mathrm{p}^*_i < 1$) and unconditional defectors. As defectors we always set the $D$ individuals with the highest values of $d/z$. This is the group of individuals for which the social dilemma requires a larger $\alpha$ as a means to disappear, if there is a positive probability for reaching a public goods state.

We find that, among the random graph models, the ER graph model requires the lowest efficiency (benefit to cost ratio) for the cooperation to persist (i.e. not to be extinct), followed by the RR graph. The BA graph presents itself as the topology where extinction of cooperation is more probable (compared with ER and RR graphs), for the same average graph degree. We remark that this observation is independent on the choice for $\varepsilon$. In contrast, it can be noticed that there is quite a colorful discrepancy with respect to the threshold above which full (i.e. unconditional) cooperation begins. In this aspect, when donation is the only natural process, the RR graph is the most supportive to cooperation, followed by the ER and the BA graph. However, when public goods is included in the possible natural states, it seems that when around half of the population is constituted by defectors the BA graph performs better in promoting unconditional cooperation, while the ER graph is the best promoter of unconditional cooperation when the majority of the population are defectors. 

Similarly to the scenario without defectors, in the case with defectors the findings can also be attributed to the distribution of the index $s$. Concretely, the inclusion of defectors can be described as an internal force which ultimately decreases the value of $s$. From this point of view, the extinction of cooperation is related to the right tail of the distribution since then one faces instances where the necessary condition for existence of cooperation is not fulfilled. As discussed previously, the ER graph has the largest tail for the distribution of the index $s$ among the studied random graphs, and therefore is most robust to extinction of cooperation when defectors are included. On the other hand, it seems considerably more delicate to precisely quantify the differences between the random graph models with respect to the existence of full unconditional cooperation as function of the distribution of the index $s$, as the results depend on the specific choice of the individuals which are selected as defectors. Nonetheless, one can notice that, as the number of defectors increases the distribution of $s$ is driven towards lower values and thus the threshold for full cooperation increases. Evidently this has lowest effect on the RR graph since the distribution of $s$ in that case concentrates around one value, i.e. there are no tails. Hence the observed lower thresholds for full unconditional cooperation in the sole donation case.  We note that, in general, this explanation does not hold when provision of public goods is a possible state in the global system since then, as shown previously, for each individual a critical point exists after which cooperation is no longer a social dilemma. For instance, the value of this critical point in the RR graph is independent of the selected individual (its minimum for each graph and structure in the figure is denoted as a black vertical line). From the figure, it appears that the ER graph requires the lowest efficiency in order to have a nonempty set of individuals for which the social dilemma disappears, followed by the BA graph. A direct corollary of this is the observed lower threshold for full unconditional cooperation in these two graphs, compared to the RR graph, when a large fraction of the population behaves as defectors.

\newpage
\section{The role of multiplex network structure in cooperation \\ through generalized reciprocity \label{sec:two}}

\subsection{Introduction}

While the model developed in the previous chapter sheds valuable insight on the role that network topology plays in promoting cooperation, it addresses only interactions on networks that are of one ``dimension'', ignoring possible multidimensional phenomena, i.e. multiplex network structures. This is an obvious drawback since real-life networks often exhibit heterogeneous properties within the edge structure that are of fundamental value to the phenomena present in the system~\cite{Kivela-2014}. For instance, in social network analyses the patterning and interweaving of different types of relationships are needed to describe and characterize social structures~\cite{Boorman-1976,White-1976}. In telecommunication networks, where control of the level of cooperation displayed by the nodes is needed to achieve efficiency~\cite{gajduk2014energy,chu2010opportunistic}, the physical edges are often ``sliced'' into multiple parts in order to support the requirement of different devices~\cite{Sherwood-2010,Nikaein-2015}. Even genetic and protein relations between organisms constructed in multiple ways are crucial for the analysis of their cooperative bindings~\cite{pandit2013genome,Stark-2006,DeDomenico-2015}. Another example is cooperation ecological systems where species interact in various ways~\cite{pilosof2017multilayer}.

To this end, here we extend the model introduced in the previous chapter to account for a multiplex network structure, with the aim to characterize the network cooperation dynamics under the assumption of a state-based behavioral mechanism rooted in generalized reciprocity. 
In our model the dimensions act as platforms which facilitate transactions between active members. The activity of the individuals is modeled by constraining their presence to one dimension per round, and by making them able to answer only to requests from that same dimension. This assumption is consistent with the random walk models on multiplex networks~\cite{de2014navigability}, and is justified in systems where the round duration is very short and/or when individuals have limited interaction capacities. The resulting mechanism, while preventing exploitation by other individuals, exhibits additional features that act as promoters of cooperation in a multiplex network structure. Specifically, by allowing for heterogeneous benefits and costs (i.e. different parameter values across different dimensions), we show that cooperation can survive in the observed dimension even if the cost exceeds the benefit, as long as there is another dimension which acts as a support (having benefit-to-cost ratio larger than one). This essential characteristic of the new model comes in contrast to one-dimensional networks where the benefit being larger than the cost is a prerequisite for cooperation. In particular, in a one dimensional network a benefit to cost ratio less than one implies that the cooperative individual has to carry a larger cost than the benefit the other individual receives, therefore it may be said that cooperation reduces the overall social welfare. In a multiplex network, we argue that this decrease in social welfare in a observed dimension is compensated by a large enough benefit to cost ratio in another dimension. Moreover, by introducing simple dynamics for the probability that an individual is present in a certain dimension, we show that, under a behavioral model based on generalized reciprocity, the cooperative contributions effectively concentrate to the dimension where most of their cooperative neighbors are also present. Based on these observations, we discuss connections to reinforcement learning, in particular to the model of Roth and Erev~\cite{roth1995learning} and extensions therein~\cite{camerer1999experience}.

\subsection{Model}
\label{sec:model}

\subsubsection{Network interactions}
\label{sec:mplexcoop_Model}

We consider a population of $N$ individuals whose relations are modeled as a connected multiplex network, defined as the triplet $\mathcal{G} \left( \mathcal{N}, \mathcal{E}, \mathcal{L} \right)$, where $\mathcal{N}$ (the set of nodes) corresponds to the set of individuals, $\mathcal{E} \subseteq \mathcal{N} \times \mathcal{N}$ is the set of edges that describes the relationships between pairs of individuals, and $\mathcal{L}$ is the set of $L$ properties that can be attributed to the edges and which define the dimensions of the network. Formally, a dimension can be defined as the graph $\mathcal{G}^{\left[l\right]} \left( \mathcal{N}, \mathcal{E}^{\left[l\right]} \right)$ in which $\mathcal{E}^{\left[l\right]}$ is the subset of $\mathcal{E}$ having the property $ l \in \mathcal{L}$. Each dimension is given via an $N \times N$ adjacency matrix $\mathbf{A}^{\left[l\right]}$, where the $ij$-th entry $A^{\left[l\right]}_{ij} \in \{0,1\}$ between pairs of individuals $i, j \in \mathcal{N}$ ($ A^{\left[l\right]}_{ij}=1$ indicating neighborhood relation, i.e. $\left(i, j \right) \in \mathcal{E}^{\left[l\right]}$). 

The interactions between the individuals are modeled as follows: in each round $t$, each individual $i$: 
\begin{enumerate}
 \item randomly chooses a dimension $l$ where it will be present in that round;
 \item sends a cooperation request to a randomly (on uniform) chosen individual $j$ from its neighborhood in the $l$-th dimension, $j\in\mathcal{N}^{\left[l\right]}_i$;
\item upon selection, if individual $j$ is present in the the $l$-th dimension in round $t$, it receives the request and cooperates with probability $\mathrm{p}_j(t)$ representing the individual's internal cooperative state at round $t$. The update rule for $\mathrm{p}_j(t)$ is the same as the one used in the previous chapter;
\item When cooperating, individual $j$ pays a cost $c^{\left[l\right]} > 0$ for individual $i$ to receive a benefit $b^{\left[l\right]} > 0$.
\end{enumerate}
Given this interaction model, the random payoff of individual $i$ at round $t$ may be characterized as
\begin{align}
\mathrm{y}_i(t)&=\sum_l v^{\left[l\right]}_{i}(t) \left[ b^{\left[l\right]} v^{\left[l\right]}_{j}(t) \mathrm{x}_{j}(t) -c^{\left[l\right]} \mathrm{x}_i(t)\sum_{k\in\mathcal{N}^{\left[l\right]}_i} \rho^{\left[l\right]}_{k}(t) v^{\left[l\right]}_{k}(t) \right].
\label{eq:mplexcoop_payoff}
\end{align}
In~(\ref{eq:mplexcoop_payoff}), $v^{\left[l\right]}_{i}(t)$ is the $i$-th outcome of an $L$ dimensional categorical variable parametrized by $B^{\left[l\right]}_{i}(t)$, which itself is a random variable describing the probability that $i$ is present in layer $l$ in round $t$. The selected index from the neighborhood of $i$ is a random variable uniformly distributed on the set $\mathcal{N}^{\left[l\right]}_i$, $j\sim \mathrm{U}(\mathcal{N}^{\left[l\right]}_i)$; $\mathrm{x}_j(t)$, $j=1,\ldots, N$, are Bernoulli random variables, each with parameter $\mathrm{p}_j(t)$; $\rho^{\left[l\right]}_k$ is a Bernoulli random variable with parameter $1/d^{\left[l\right]}_k$, where $d^{\left[l\right]}_k$ is the degree of individual $k$ in dimension $l$, $d^{\left[l\right]}_k=\sum_h A^{\left[l\right]}_{kh}$; the term $\sum_{k\in\mathcal{N}^{\left[l\right]}_i} \rho^{\left[l\right]}_{k}(t)$ captures the random number of individuals (neighbors of $i$ in $l$) which send a cooperative request to $i$ in dimension $l$ during round $t$. 

\subsubsection{Deterministic approximation} 
\label{sec:mplexcoop_Deterministic_Model}
We approximate the stochastic model~(\ref{eq:mplexcoop_payoff}) by a deterministic model in which the random variables are substituted with their respective expectations
\begin{align}
\mathrm{y}_i(t)&=\sum_l B^{\left[l\right]}_{i}(t) \left[ b^{\left[l\right]} \sum_j \frac{A^{\left[l\right]}_{ij}}{d^{\left[l\right]}_i} B^{\left[l\right]}_{j}(t) \mathrm{p}_{j}(t) - c^{\left[l\right]} z^{\left[l\right]}_i(t)  \mathrm{p}_{i}(t) \right].
 \label{eq:deterministic_mplexcoop_compact}
\end{align}
The term $z^{\left[l\right]}_i(t)$ in (\ref{eq:deterministic_mplexcoop_compact}) is defined as
\begin{align}
z^{\left[l\right]}_i(t)=\sum_j \frac{A^{\left[l\right]}_{ji}}{d^{\left[l\right]}_j} B^{\left[l\right]}_{j}(t),
\end{align} 
is the temporal extension of the neighborhood importance index discussed in~\cite{Utkovski-2017}, in dimension $l$. This quantity acts as a local centrality measure of an individual, with individual $i$ being more ``important'' in the addressed dimension if it has many neighbors which are at the same time also present in that dimension, and the neighbors themselves have few neighbors. In our model of interactions, this individual would be called upon rather frequently in the studied dimension.   

\begin{figure*}[t]\centering
 \begin{adjustwidth}{0.9in}{0in}
\includegraphics[width=13cm]{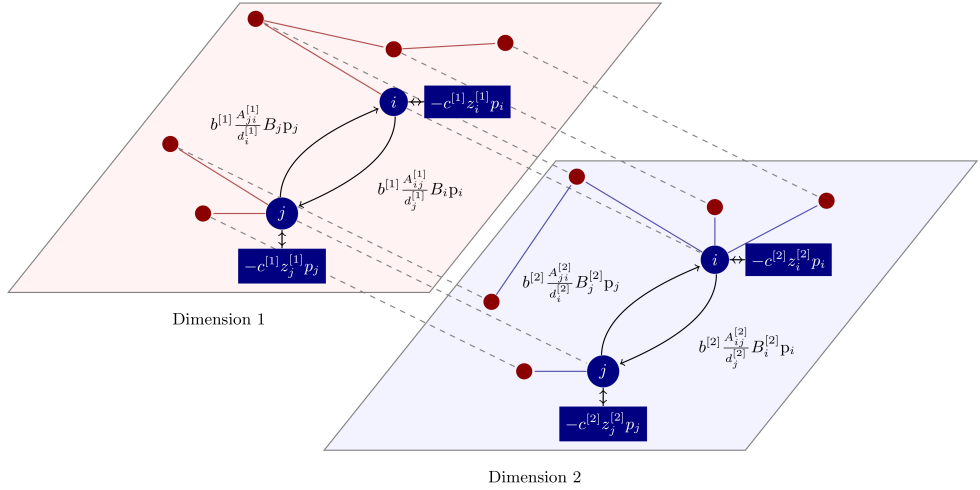}
\end{adjustwidth}
\caption{An example for the deterministic interactions on a two dimensional network between two individuals $i$ and $j$ (colored in dark blue). For illustrative purposes we exclude the round notation. Filled colored edges indicate neighborhood relation in the corresponding dimension, whereas dashed lines are links to the same individual in the other dimension.}
\label{fig:mplex_coop_eg}
\end{figure*}

The motivation to use the deterministic model~(\ref{eq:deterministic_mplexcoop_compact}) is that it captures the long-term behavior of the stochastic model, i.e. provides reliable approximation of its steady state behavior. Numerical simulations of the stochastic model in the long run suggest that the stochastic variables can indeed be approximated by their respective expectations, without affecting the long-term network behavior, thus justifying the approximation. In Fig.~\ref{fig:mplex_coop_eg} we display an example for the deterministic interactions between two individuals that are placed on a two-dimensional network.

We also define the aggregate time-dependent neighborhood importance index $Z_i(t)$ of individual $i$ as
\begin{align}
Z_i(t) &= \frac{\sum_l B^{\left[l\right]}_{i}(t) z^{\left[l\right]}_i(t)}{\sum_l B^{\left[l\right]}_{i}(t) \sum_j \frac{A^{\left[l\right]}_{ij}}{d^{\left[l\right]}_i}B^{\left[l\right]}_{j}(t)}.
\end{align}
As we will see in more detail in what follows, the probability distribution of the quantity $Z(t)$ across the individuals crucially determines the global cooperative behavior in the network. This quantity, which is a form of aggregated centrality measure in the multiplex network setting, critically reflects the role of the network topology on the cooperation dynamics in our interaction model. 

With the above, we write~(\ref{eq:deterministic_mplexcoop_compact}) in a more compact (vector) form as  
\begin{align*}
\mathbf{y}(t) = \mathbf{\Theta}(t)\cdot\mathbf{p}(t), 
\end{align*}
where $\Theta_{ii}(t)= -\sum_l c^{\left[l\right]} B^{\left[l\right]}_{i}(t) z^{\left[l\right]}_i(t)$, and $\Theta_{ij} (t) = \sum^{\left[l\right]} b^{\left[l\right]} B^{\left[l\right]}_{i}(t) \frac{A^{\left[l\right]}_{ij}}{d^{\left[l\right]}_i} B^{\left[l\right]}_{j}(t)$, for $i\neq j$.

\subsection{Results}
\label{sec:Results}

\subsubsection{Analytical properties of the model} 
\label{sec:analytical_properties}
ere we characterize the main properties of the model in steady state. Hereby, we distinguish between two types of results: 1) results with homogeneous parameters, defined as identical benefit and costs across network dimensions, $b^{\left[l\right]} = b$ and $c^{\left[l\right]} = c$ for all $l \in \mathcal{L}$; and 2) results with heterogeneous parameters, i.e. the more general case when we allow for different values for the benefits and costs across network dimensions. We remark that the proofs for the properties follow directly from applying the framework presented in references~\cite{stojkoski2018cooperation,Utkovski-2017}, nevertheless for concreteness in the exposition we present them.

The following property holds in general. 

\textbf{1.~Robustness~to~exploitation.} In steady state, the individuals may be attributed to two disjoint sets, $\mathcal{W} = \left\{ w \in \mathcal{N} : \mathrm{y}^*_w=0  \right\}$ and $\mathcal{S} = \left\{ s \in \mathcal{N} : \mathrm{y}^*_s>0  \right\}$, based on the steady state payoff $\mathrm{y}^*_i$. The individuals in $\mathcal{S}$, which we refer to as ``strong individuals'', are characterized by $\mathrm{p}^*_i=1$, while the individuals in $\mathcal{W}$, called ``weak'' may take both values $\mathrm{p}^*_i=1$ and $\mathrm{p}^*_i<1$, depending on the network parameters. Hence, there are two sets of relations that have to be satisfied 
\begin{align}
0& =\sum_l B^{\left[l\right]*}_{i} \left[ b^{\left[l\right]} \sum_j \frac{A^{\left[l\right]}_{ij}}{d^{\left[l\right]}_i} B^{\left[l\right]}*_{j} \mathrm{p}_{j}^* - c_l z^{\left[l\right]*}_i  \mathrm{p}_{i}^* \right], \:\: i\in \mathcal{W}\nonumber\\
\mathrm{y}^*_i &= \sum_l B^{\left[l\right]*}_{i} \left[ b^{\left[l\right]*} \sum_j \frac{A^{\left[l\right]}_{ij}}{d^{\left[l\right]}_i} B^{\left[l\right]*}_{j} \mathrm{p}_{j}^* - c^{\left[l\right]} z^{\left[l\right]*}_i  \mathrm{p}_{i}^* \right], \:\: i\in \mathcal{S}.
\label{eq:optimization_model_simplified}
\end{align}  
Note that the sets $\mathcal{W}, \mathcal{S}$, the steady state values $\mathrm{p}_i^*,\:i\in\mathcal{W}$ and $B^{\left[l\right]*}_{j},\:i\in\mathcal{N},\:l\in\mathcal{L}$ and the constants $\mathrm{y}^*_i, \:i\in \mathcal{S}$ are unknown. 

The next properties hold only for a multiplex network with homogeneous parameters. We relax this assumption in the numerical analysis performed in the following sections.

\textbf{2.~Existence~of~cooperation.} A necessary condition for existence of cooperators (individuals with $\mathrm{p}^*_i>0$) is $b/c \geq 1$.

\textbf{3.~Promotion~of~cooperation.} When $b/c>1$, we observe the steady state probabilities are strictly greater than $0$, $\mathrm{p}_i^*>0$ for all $i\in \mathcal{N}$.

\textbf{4.~Sufficient~condition~for~existence~of~strong~individuals.} When $b/c>1$, there is always at least one strong individual in the network.

\textbf{5.~Necessary~condition~for~the~existence~of~strong~individuals.} A necessary condition for existence of strong individuals, (individuals with $\mathrm{p}^*_i = 1$), is $Z^*_i \leq b/c$.

\textbf{6.~Full~network~cooperation.} The condition $b/c \geq Z^*_{max}$, where $Z^*_{max}$ is the largest neighborhood importance index in the graph, $Z^*_{max} = \max_i Z^*_i$, is both necessary and sufficient for all individuals to be strong.

\subsubsection{The role of heterogeneous parameters}
\label{sec:heter-pars}

We continue the analysis by relaxing the assumption of homogeneous parameters, and consider the situation where each dimension $l$ has its own benefit $b^{\left[l\right]}$ and cost $c^{\left[l\right]}$. Since our goal is to examine the effect of heterogeneous parameters, we develop a null model in which the probability for dimension presence is uniform in every round.

For this case, we compare two different multiplex networks each composed of two dimensions (the results can be easily generalized to networks with more dimensions). In particular, the first multiplex network type represents a natural generalization of the Erdos-Renyi (ER) random graph, while the second is a multiplex version of the Barabasi-Albert (BA) preferential attachment graph. In an ER random graph an edge between two individuals has a fixed probability of being present, independently of the other edges. As a consequence the degree follows a Poisson distribution. On the other hand, a BA graph is constructed by a dynamical process in which in each step a new individual is introduced, and this individual makes connections to other individuals that are already in the graph with probability proportional to their degree, thus ending up with a power law degree distribution. More details about the algorithms used for generating these multiplex networks can be read in references~\cite{bianconi2013statistical,nicosia2013growing,nicosia2014nonlinear}.

In what follows, we will consider networks which consist of 100 individuals, and where the average degree in each dimension is 8. Since the correlations between the edges in different dimensions should play a prominent role in determining the steady state level of cooperation we are going to study three different scenarios. In the ER graph, we will examine the possibility of overlapping edges, i.e. situations i) when there is no edge overlap in different dimensions, ii) when half of the edges overlap, and iii) when all edges overlap. In the BA graph we study the cases where the degree correlation $\rho$ generated through preferential attachment is i) negatively correlated, ii) there is no correlation, and iii) is positively correlated between the dimensions. Formally we measure the correlation between the degrees in separate dimensions through the Pearson correlation coefficient, i.e
\begin{align}
\rho_{12} &= \frac{\sum_i \left( d^{\left[1\right]}_i - \langle d^{\left[1\right]}\rangle \right) \left( d^{\left[2\right]}_i - \langle d^{\left[2\right]}\rangle \right) }{  \sqrt{\sum_i \left( d^{\left[1\right]}_i - \langle d^{\left[1\right]}\rangle \right)^2} \sqrt{\sum_i \left( d^{\left[2\right]}_i - \langle d^{\left[2\right]}\rangle \right)^2}}.
\label{eq:pearson_correlation}
\end{align}
In equation (\ref{eq:pearson_correlation}) $\langle d^{\left[l\right]}\rangle$ denotes the average degree in dimension $l$.

\begin{figure*}[t!]
\includegraphics[width=12cm]{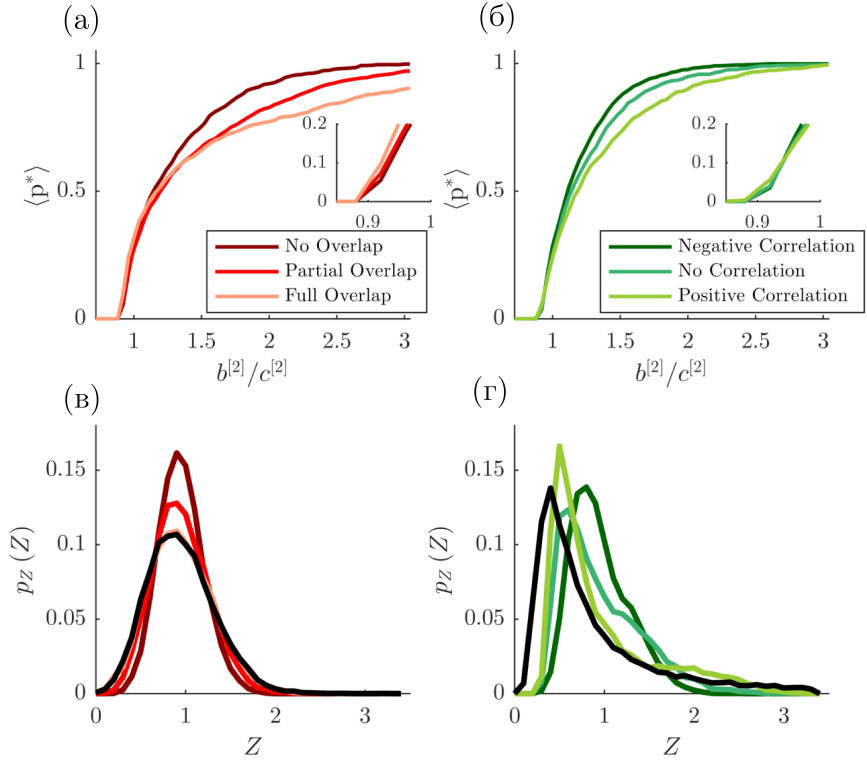}
\caption{\textbf{(a-b)} Fraction of strong individuals $\sigma$ as a function of the benefit to cost ratio in the second dimension, $b^{\left[2\right]}/c^{\left[2\right]}$, for a sample of the random graphs that we study, while $b^{\left[1\right]}/c^{\left[1\right]} = 1.08$. \textbf{(a)} ER multiplex networks with uniform update. \textbf{(b)} BA multiplex networks with uniform update. \textbf{(c-d)} Probability density function for the steady state $Z$ index for the same random graphs averaged across 1000 realizations and the corresponding one-dimensional graphs (in black). \textbf{(c)} ER multiplex networks. \textbf{(d)} BA multiplex networks. \textbf{(c-d)} The black curve illustrates the distribution of $Z$ in the one-dimensional representation of the corresponding random graph. All networks have 100 individuals and average degree 8.  }
\label{fig:heterogen-pars}
\end{figure*}

The results are depicted in Fig.~\ref{fig:heterogen-pars}. Panels~(a)-(b) respectively show the evolution of the fraction of strong individuals for the multiplex network generated by the ER and BA multiplex networks while the benefit to cost ratio in the second dimension, $b^{\left[2\right]} / c^{\left[2\right]}$ is varied, whereas the benefit to cost ratio in the first dimension is kept constant ($b^{\left[1\right]} / c^{\left[1\right]} = 1.08$).  
In the case of ER networks, it can be noticed that full edge overlap is better for the level of cooperation displayed when $b^{\left[2\right]} / c^{\left[2\right]}$ is low (see the inset plot of Fig.~\ref{fig:heterogen-pars}~(a)), while no edge overlap promotes more cooperation for higher values of the benefit to cost ratio in the second dimension. Similarly, positive degree correlation in the BA networks leads to higher fraction of strong individuals for small benefit to cost ratios(inset plot of Fig.~\ref{fig:heterogen-pars}~(a)), while negative degree correlation is better for supporting cooperation when $b^{\left[2\right]} / c^{\left[2\right]}$ is high.

We point out that due to the uniform probability of dimension update there is a symmetric relationship between changes in the benefit to cost ratios in both dimensions and the fraction of strong individuals, i.e. a change in $b^{\left[l\right]} / c^{\left[l\right]}$ in one dimension has the same effect as a change in the other dimension. Any slight adjustment in the probabilities only changes the symmetry towards one of the dimensions, which means that the results should not be gradually altered. Hence, we can use the properties derived in the previous section as a starting point for the inspection of the results.

This reduces the analysis to studying the steady state distribution of the index $Z$. We note that, according to our interaction model, higher values of the index imply more frequent cooperation requests, and hence, lower incentives for cooperation. For this purpose in panels~(c)-(d) of Fig.~\ref{fig:heterogen-pars} we plot the typical probability density function (PDF) for the $Z$ index for the same random multiplex networks where the dimension presence is given by a uniform probability by averaging the index across 1000 network realizations. There, in black, we also depict the PDF of $Z$ for the one dimensional ER and BA networks\footnote{Note that the full overlap multiplex ER graph coincides with the one-dimensional representation.}. For the ER graphs, it can be seen that the exclusion of overlapping edges effectively increases the mode of the distribution. Since the average level of cooperation at the point where cooperation begins to exist is determined by the left tail of the distribution, the networks with lower modes, and thus larger left tails, should be able to promote more cooperation. Contrastingly, the fatness of the right tail leads to higher thresholds for displaying full network cooperation. In a similar fashion, we notice that by decreasing the correlation between the degrees in the multiplex BA graph, the right tail of $Z$ decreases, and therefore the lower threshold for full cooperation.

Finally, in the inset plots of  Fig.~\ref{fig:heterogen-pars} (a)-(b) we observe that the inclusion of a second dimension leads to significantly lower threshold for existence of cooperation in the system (which in the one dimensional case is $b / c > 1$~\cite{Utkovski-2017,stojkoski2018cooperation}). This implies that the other dimension acts as a support for existence of cooperation even if the original dimension does not allow it. This is a result of the fact that the negative payoffs from the original dimension are compensated with positive payoffs from the supporting dimension. If at least one individual receives higher steady state payoff from the supporting dimension than the loss in the original, then cooperation will persist. This is an important implication to the emergence of cooperation in systems where all dimensions of the network can not be observed and the environment is not suited for cooperation due to not increasing the social welfare, while the phenomena is still detected.

\subsubsection{The role of dynamics in the dimension update rule}
\label{sec:dimension_update}

Predetermined presence is a plausible assumption for systems where the flow between dimensions is constrained and individuals are not allowed to develop beliefs about which dimensions generate higher payoffs to them. A more realistic case would be to allow for dynamics in the probability that individual $i$ is present in dimension $l$ in round $t$. While this can be modeled by introducing Markov transition rates for moving from one dimension to another, or even adding memory rates to the movement based on the experience in the previous rounds, here we consider a simpler update rooted in the same generalized reciprocity rule that was used for the internal state update. 

Concretely, we consider an update based on the accumulated payoff in the dimension, 
\begin{equation}
\mathrm{Y}^{\left[l\right]}_{i}(t)=\mathrm{Y}^{\left[l\right]}_{i}(t-1)+\mathrm{y}^{\left[l\right]}_{i}(t),
\label{eq:payoff_propensity}
\end{equation}
with $\mathrm{Y}^{\left[l\right]}_i(0)$ being the initial condition and $\mathrm{y}^{\left[l\right]}_{i}(0)=0$. In our model the updated probability of presence of individual $i$ in dimension $l$ is given by the softmax function
\begin{align}
B^{\left[l\right]}_{i}(t+1) = \frac{\exp\left( \mathrm{Y}^{\left[l\right]}_{i}(t) \right) }{ \sum_{m} \exp\left( \mathrm{Y}^{\left[m\right]}_{i}(t) \right)}.
\label{eq:dimension_update}
\end{align}
We remark that the described rule is similar in spirit to the famous Roth-Erev reinforcement learning algorithm for strategies in extensive form games~\cite{roth1995learning}. In fact, based on (\ref{eq:payoff_propensity}) and (\ref{eq:dimension_update}), an analogy with more general reinforcement learning models can be established~\cite{camerer1999experience,jost2014reinforcement}. The connection is provided by interpreting the act of presence of individual $i$ in dimension $l$ of the multiplex network as a selection of a strategy $S_l$ (from a set of $L$ preselected strategies), where the strategy selection is applied with probability $B^{\left[l\right]}_{i}$. In this context, the total payoff $\mathrm{Y}^{\left[l\right]}_{i}$ in the network dimension $l$ is analogous to the propensity to play strategy $S_l$. We note that the here applied  rule~(\ref{eq:dimension_update}) is different from the one introduced in the original Roth-Erev learning model, according to which strategy $l$ is selected with probability $B^{\left[l\right]}_{i}(t+1) = \frac{\mathrm{Y}^{\left[l\right]}_{i}(t)}{ \sum_{m} \mathrm{Y}^{\left[m\right]}_{i}(t)}$. Specifically, it can be considered as a special case of a more general reinforcement learning scheme in which the
probabilities for players to choose certain actions are taken
from a general Gibbs-Boltzmann distribution
\begin{equation}
B^{\left[l\right]}_{i}(t+1) = \frac{\exp\left( \lambda \cdot \mathrm{Y}^{\left[l\right]}_{i}(t) \right) }{ \sum_{m} \exp\left( \lambda \cdot \mathrm{Y}^{\left[m\right]}_{i}(t) \right)}.
\label{eq:Gibbs_Boltzmann}
\end{equation}
In~(\ref{eq:Gibbs_Boltzmann}) $\lambda$ plays the role of ``inverse temperature'' in statistical physics, and captures the trade-off between exploitation ($\lambda=\infty$), i.e. greedy
learning in which only the action with the highest propensity is taken, and exploration ($\lambda=0$), meaning that all actions are equally probable. In many reinforcement learning problems, the key is to find a value of $\lambda$ that achieves a reasonable trade-off between exploitation and exploration in the model of question. In our model, the selection of this parameter would critically determine the transient and evolutionary behavior in the system when, for example, some dimensions are erased. We expect that in that case selecting the parameter $\lambda$ towards the ``exploration mode'' would provide certain robustness to such events. The exact quantification of these effects under this scenario, together with the role of the network topology, is out of the scope of this doctoral thesis. However, it represents an interesting direction for future work.

The advantage of the suggested update is that it can be very easily implemented since the individual only needs to know the probability $B^{\left[l\right]}_{i}(t)$ in the current round for all dimensions and the received payoff $\mathrm{y}^{\left[l\right]}_{i}(t)$ from them.

We point out that the steady state solution of equation~(\ref{eq:dimension_update}) is not defined if more than one $\mathrm{Y}^{\left[l\right]}_{i}$ tends to infinity. This never happens as long as each individual experiences different dynamics when the dimensions are considered as separate networks. Another thing worth emphasizing is that the resulting system has $\left( L - 1 \right)N$ degrees of freedom, and, hence, complex behavior is unavoidable. Therefore, in the comparative statics as starting points for the probability that an individual is present in a certain dimension we consider real numbers whose values are comparable to the steady state of the preceding benefit to cost ratios. In the beginning, at the lowest benefit to cost ratios, the starting point is set to be equal among all individuals and dimensions. 

\begin{figure*}[t!]
\includegraphics[width=12cm]{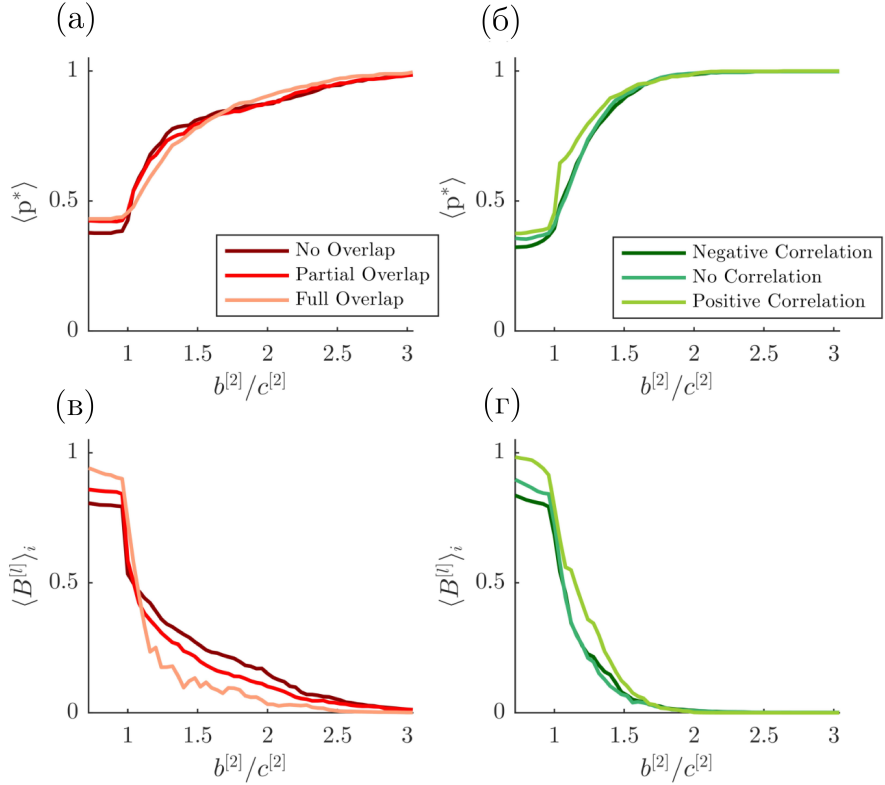}
%\vspace{-1.2cm}
\caption{\textbf{(a-b)} Fraction of strong individuals $\sigma$ as a function of the benefit to cost ratio in the second dimension, $b^{\left[2\right]}/c^{\left[2\right]}$, for a sample of the random graphs that we study, while $b^{\left[1\right]}/c^{\left[1\right]} = 1.08$. \textbf{(a)} ER multiplex network with generalized reciprocity update. \textbf{(b)} BA multiplex network with generalized reciprocity update. \textbf{(c-d)} For the same graphs, the fraction of individuals $\langle B^{\left[l\right]*} \rangle$ which in the steady state are present in the first dimension.}
\label{fig:dimension-update}
\end{figure*}

The results for the same networks as in the previous section are shown in Fig.~\ref{fig:dimension-update}. Panels~(a)-(b) depict the fraction of strong individuals as a function of the benefit to cost ratios. On the one hand, we observe that the global level of cooperation displayed for low $b^{\left[2\right]} / c^{\left[2\right]}$ ratios is increased by a large amount when compared to the predetermined probability for dimension presence. On the other hand, we notice that for larger benefit to cost ratios, the overall level of displayed cooperation is not consistent in terms of performance. More precisely, there are situations in which it is decreased (e.g. the no-overlap ER network), and there are situations in which it is increased (e.g. the positive correlation BA network) when compared to predetermined presence.

This aggregate behavior can be explained by looking at (c)-(d) of Fig.~\ref{fig:dimension-update}, where we display the fraction of individuals which in steady state are present in the first dimension as a function of the same parameters. Obviously, the dimension in which the individuals are always present in steady state is not always the same, i.e. it is dispersed among the individuals depending on the network topology and parameters. This is a key feature of the model since it implies that the dimension update rule forces the individuals to accommodate their presence towards the dimension where they either carry the smallest burden to cooperate or where most of their cooperative neighbors are present. As such, when coupled with the generalized reciprocity state update rule (\ref{eq:update}), the dimension update rule facilitates the promotion of cooperation in the system (in the sense that it promotes the existence of individuals with $\mathrm{p}^*_i > 0$). 

It is worth mentioning that the fact that individuals may accommodate their presence to the dimension where most of their cooperative neighbors are present indicates that full unconditional cooperation is not guaranteed to be achieved in an easier fashion. In particular, the dimension update rule may force an individual in steady state to be present in a dimension which is less suited for its personal gain than some predetermined rule just because most of the other individuals are present in that dimension. Exactly this may be the cause for the lower level of cooperation exhibited in the no-overlap ER network. Due to the manner in which the edges are constructed in this graph, it may happen that some individuals to have high neighborhood importance index in the first dimension but low in the other. For these individuals, a uniform predetermined rule would imply that they would be able to generate a higher payoff by interacting in the dimension where they are less burdened. However, none of their neighbors with lower value of $Z$ would be better of if they are present in it due to lower benefit to cost ratio and, more importantly, due to being less forced to answer to a cooperation request. As a consequence, these group of individuals accommodate their presence in the second dimension, and the individuals with high neighborhood importance index in it are also forced to be present in it. 

\subsection{Conclusion}

In this chapter of the doctoral thesis we studied the cooperation dynamics under a behavioral mechanism based on generalized reciprocity, in a network consisting of multiple dimensions, each modeled by a random graph. The model, which is a generalization of the results described in the previous chapter, provides new insights on the role of the network structure on the promotion of cooperation in complex multidimensional networks. In particular, we demonstrated that a multidimensional structure may support cooperation within the individual network dimensions, even when the benefit-to-cost ratio in the considered dimensions is below the threshold required for cooperation (when observed in isolation). This observation may explain the existence of cooperation in systems where cooperative behavior is observed even though it does not increase social welfare -- a latent structure (i.e. other dimensions) may exist that acts as a support to the observed dynamics.

We also discussed the connection between the studied behavioral mechanism in the multidimensional network and reinforcement learning, by interpreting the act of presence of individuals in the dimensions of the multiplex network as a selection of a strategy (from a predefined set of strategies), where the strategy selection is applied in relation to individuals' internal state. In this context, we introduced a simple and intuitive rule for modeling the individuals' interactions in the different dimensions (i.e. their presence across dimensions). As a general observation, we found out that the cooperative contributions of the individual individuals concentrate in the dimension which is most favorable for the existence of cooperation.

An interesting direction for future work is the study of more general behavioral mechanisms in the spirit of the ``exploration vs. exploitation'' discussion in reinforcement learning. In this context, it will be valuable to study not only the steady-state (in the sense of evolutionary behavior), but also the transient dynamics in the network. Another fruitful topic is generalizing the multiplex model to other truly cooperative games that cannot be sustained in absence of other behavioral mechanism. As a final note, the model can also be used as a starting point in the examination of network formation based on generalized reciprocity, where the neighborhood of each individual in each dimension can be seen as the possible final outcome of a rewiring process that is determined by the dimension update rule.

\newpage
\section{Evolution of cooperation in networked \\ fluctuating environments \label{sec:three}}

\subsection{Introduction}

%Кооперацијата игра фундаментална улога во еволуцијата на различни нивоа од мрежни организации, од прости клеточни па се до сложени човечки интеракции~\cite{Axelrod-2006}. Сепак, природната селекција наметнува конкуренција и поради тоа кооперативното однесување се темели на појава на специфичен механизам во рамките на истражуваната мрежа~\cite{Nowak-2006five}. 

A ubiquitous, yet largely unexplored, feature in the cooperative dynamics on complex networks is the presence of a fluctuating environment, i.e. situations where the temporal evolution is strongly affected by relative movement. In such situations, fluctuations have a net-negative effect on the time-averages, although having no effect on the ensemble properties~\cite{peters2013ergodicity}. This observation, which captures the non-ergodicity of fluctuating environments, yields evolutionary behavior which essentially differs from the one observed in standard models~\cite{radicchi2018uncertainty,stollmeier2018unfair}.

In this chapter, we investigate the impact of complex networks on the cooperative dynamics in fluctuating environments, by studying the networked pooling and sharing of accumulated payoffs whose growth undergoes a multiplicative process. Noisy payoff growth is a common model for  fluctuations~\cite{stollmeier2018unfair,zheng2018environmental,cvijovic2015fate}, and pooling and sharing has been found as a robust promoter of cooperation under these circumstances~\cite{yaari2010cooperation,liebmann2017sharing,peters2015evolutionary}. While there remains the general trend that cooperation reduces fluctuations (i.e. is evolutionary advantageous), we show that the unique interplay between the non-ergodic fluctuation-generating process and the network topology may generate large discrepancies in the individual resource endowments. When present, this inequality has a negative effect on the growth rates of the individuals, hampering their evolutionary performance. Parallels can be made to current societal discussions on inequality~\cite{sciam2018}.

\subsection{Model}

Formally, we assume that the dynamics of the accumulated payoff $Y_i(t)$ of individual $i$ at time $t$ follow a geometric Brownian motion (GBM),

\begin{align}
\mathrm{d} Y_i &= Y_i \left( \mu \mathrm{d}t + \sigma \mathrm{d}W_i \right),
\label{eq:gbm}
\end{align}
with $\mu$ being the drift term, $\sigma$ the amplitude of the noise and $\mathrm{d}W_i$ is an independent Wiener increment, i.e. $W_i(t) =\int_0^t \mathrm{d}W_i$. Without the noise, i.e. $\sigma = 0$, the model is simply exponential growth at rate $\mu$. With $\sigma \neq 0$ it can be interpreted as exponential growth with a fluctuating growth rate. The advantage of modelling through GBM lies in its universality, as the process represents an attractor of more complex models that exhibit multiplicative growth~\cite{aitchison1957lognormal,redner1990random}. Its non-ergodicity is manifested in the difference between the growth rate observed in an individual trajectory and the ensemble average growth~\cite{peters2013ergodicity, peters2016evaluating}. In particular, the estimator for the growth rate of a single GBM trajectory is defined as
\begin{align}
  g_i(Y_i(t), t) &= \frac{1}{t}\log\left(\frac{Y_i(t)}{Y_i(0)}\right),
\end{align}
where $Y_i(0)$ is the initial condition. For simplicity, we assume that $Y_i(0)=1$. 

The time-average growth rate is found by letting time remove the stochasticity in the process, i.e., taking the limit as $t \to \infty$. This is the solution
\begin{align}
\lim_{t \to \infty} g_i(Y_i(t), t) &= \mu - \frac{\sigma^2}{2}.
\label{eq:time-average-growth}
\end{align}
The ensemble growth rate, on the other hand, is found by substituting $Y_i(t)$ with the average $\langle Y \rangle$ of an infinite ensemble, where $\langle \cdot \rangle$ is the averaging operation. In other words, this is done by letting the spatial dimension remove the stochasticity by averaging across all possible realizations. Mathematically, the solution is
\begin{align}
\lim_{N \to \infty} g_i(\langle Y \rangle, t) &= \mu,
\label{eq:ensemble-average-growth}
\end{align}
where $N$ is the ensemble size.

If only a single system is to be modeled, on the long run only~(\ref{eq:time-average-growth}) is observed~\cite{peters2013ergodicity}. The ensemble average growth rate~(\ref{eq:ensemble-average-growth}) is fictive, as it assumes averaging over imagined parallel universes. Hence, in reality, it is the time-average growth rate that determines the evolutionary performance of an individual GBM trajectory. Simultaneously, it provides parallels to real-life phenomena. For instance, in evolutionary game theory the time-average growth rate is the geometric mean fitness for the accumulated payoff (i.e. resources) of a particular phenotype~\cite{saether2015concept}. In economic decision theory, where wealth dynamics follows a multiplicative process, the same growth observable arises naturally as the unique utility measure~\cite{peters2016evaluating}.

From an evolutionary perspective, individuals with lower noise amplitude should exhibit higher long-run growth rates and should thus be favored. In this regard, pooling and sharing may constitute a fundamental mechanism for the evolution of cooperation in well-mixed fluctuating environments by reducing the uncertainties in future growth and, hence, bringing closer the observed growth rate to the ensemble value~\cite{yaari2010cooperation,liebmann2017sharing,peters2015evolutionary}. For GBM dynamics, this has been nicely evidenced in~\cite{peters2015evolutionary}. Concretely, the pooling and sharing mechanism can be described as follows. A mutation wires cooperation in a population of $N$ individuals whose accumulated payoffs dynamics follow a GBM trajectory. In a discretized version of~(\ref{eq:gbm}), after a period of growth, the individuals pool their accumulated payoffs and subsequently share them \textit{equally}, resulting in the following dynamics for the accumulated payoffs
\begin{align}
\mathrm{d} Y &= Y \left( \mu \mathrm{d}t + \frac{\sigma}{\sqrt{N}} \mathrm{d}W \right).
\label{eq:gbm-pooled}
\end{align}
In~(\ref{eq:gbm-pooled}) the subscript $i$ has been dropped due to the equal sharing and $\mathrm{d}W = \frac{1}{\sqrt{N}}\sum_i \mathrm{d}W_i$ represents the pooled Wiener increment.

Evidently, equation~(\ref{eq:gbm-pooled}) is a GBM with an amplitude of $\sigma/\sqrt{N}$, thus yielding a time-average growth of
\begin{align}
g_i(Y_i(t),t) = \mu - \frac{\sigma^2}{2}\frac{1}{N}.
\end{align}

Notice that as the number of cooperating individuals increases, the time-average growth rate converges to the ensemble average growth. This implies that in finite populations, the inclusion of new individuals always produces a net performance gain. As a result, the pooling and sharing mechanism has been linked with the evolution of cooperation at the lowest possible biological level -- the transition towards multicellularity, where a species of non-cooperating single cells mutates to a species of multicellular organisms, sharing nutrients through common membranes~\cite{short2006flows,roper2013cooperatively}. Similar analogy may hold at higher levels of intelligence. As an illustration, consider situations where individuals join a community-supported agriculture to exchange their produced goods for a fixed basket of products, thereby reducing the risks in farming~\cite{adam2006community}. Another example are nations joining unions to assure sustainable economic growth through common goals~\cite{sapir2004agenda}. 

\begin{figure}[t!]
\includegraphics[width=8.6cm]{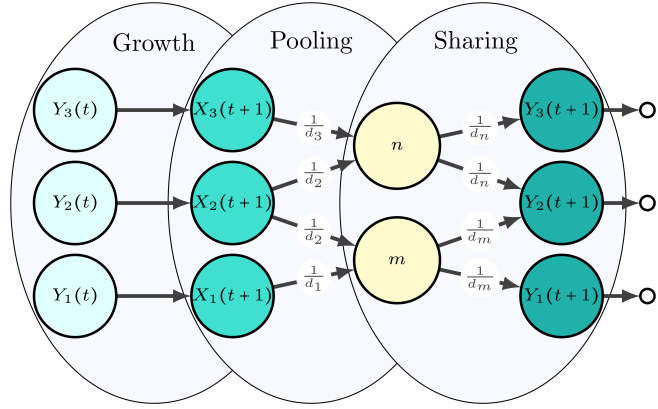}
\caption{\textbf{Networked pooling and sharing of accumulated payoffs.} The accumulated payoffs of three individuals grow according to~(\ref{eq:gbm}) and after that they are pooled in $n$ and $m$. Finally, the pools distribute the pooled accumulated payoffs equally among its participants. For visualization purposes we set $\mathrm{d}t = 1$.  \label{fig:model}}
\end{figure}

However, real-life interactions between individuals are seldom realized in a well-mixed structure, and are instead driven by a complex network of contacts~\cite{allen2017evolutionary}. To model this situation, we characterize each individual $i$ with participation in $d_i$ pools. In a discretized version of the model, each round $t$ begins with a \textit{growth phase} where the accumulated payoff of $i$ grows to $\bar{Y}_i(t+\mathrm{d}t)$. The growth phase is followed by a \textit{cooperation phase} where each individual pools an equal fraction of its accumulated payoff in each of the pools it belongs to. Afterwards, each pool returns an equal fraction of the pooled accumulated payoffs to each individual. The resulting mechanism is illustrated in Fig.~\ref{fig:model}.

The interaction structure is modeled by a connected bipartite random graph $\mathbf{B}$ between finite sets $\mathcal{N}$ of $N$ individuals and $\mathcal{M}$ of $M$ pools, with binary edge variables $B_{im} \in \left\{0,1 \right\}$ between pairs of individuals $i\in \mathcal{N}$ and pools $m\in \mathcal{M}$ ($\mathrm{B}_{im} = 1$, indicating participation of $i$ in pool $m$). The bipartite representation offers a principled way of capturing wider information regarding the group composition and network interactions~\cite{perc2013evolutionary}. In this regard, the model can be related to games of public goods played on networks~\cite{perc2013evolutionary,santos2008social,stojkoski2018cooperation}.

%In a discretized version, each round begins with a growth phase where the resources of $i$ grow analogously to~(\ref{eq:gbm}). After this, there is a \textit{cooperation phase} in which each individual $i$ pools a fraction of its resources in each of the pool it belongs, proportional to the entries of $\mathrm{B}_{im}$ and normalized for its degree $d_i$, where $d_k = \sum_l B_{kl}$. In the same manner, each pool $m$ returns a fraction $\mathrm{B}_{im} / d_m$ of the pooled resources in it to each individual $i$. The networked pooling and sharing of geometric Brownian motions is illustrated in Fig.~\ref{fig:model}.

By setting $\mathrm{d}t \to 0$, the dynamics can be explained as
\begin{align}
\mathrm{d} Y_i &= \left[\sum_j^N \mathrm{A}_{ij} Y_j - Y_i\right] \mathrm{d}t + \sum_j^N \mathrm{A}_{ij} Y_j \left( \mu \mathrm{d}t+ \sigma \mathrm{d}W_j  \right),
\label{eq:network-gbm}
\end{align}
where $\mathbf{A}$ represents a transition matrix of the network with entries $\mathrm{A}_{ij} = \sum_m^M  \frac{B_{im}}{d_m} \frac{B_{jm}}{d_j}$ determining the total allocated resources from individual $j$ to individual $i$. Equation~(\ref{eq:network-gbm}) resembles the Bouchaud--Mezard wealth reallocation model~\cite{bouchaud2000wealth,garlaschelli2008effects,berman2017empirical,ichinomiya2012wealth}, with the main difference that now the reallocation happens \textit{after} the growth phase. 

\subsection{Results}
\subsubsection{Time-average growth rate}
For tractability, we proceed by examining a discrete version of equation~(\ref{eq:network-gbm}),
\begin{align}
Y_i(t+1) = \sum_j A_{ij} Y_j(t) \left[ 1 + \mu + \sigma \varepsilon_j(t) \right],
\label{eq:network-gbm-discrete}
\end{align}
where $\varepsilon_j(t)$ is a random variable following the standard Gaussian distribution, and utilize a mean-field approach. For this purpose, we define two variables. First, the grown payoffs of each individual $i$ are given as 
\begin{align*}
X_i(t+1) &= Y_i(t) \left[ 1 + \mu + \sigma \varepsilon_i(t) \right],  
\end{align*}
For large $t$ the time-average growth rate of this variable should be the same as $g_i(Y_i(t),t)$ as its value will be dominated by $Y_i(t)$.
Second, we define the mean-field around individual $i$ as the average grown accumulated payoffs of each of its neighbors weighted by their contributions to $i$, i.e.,
\begin{align*}
\langle X_i \rangle = \frac{\sum_j A_{ij}X_j}{ \sum_j A_{ij}}.  
\end{align*}
By combining the last two equations and adapting the time scale such that $\Delta t = 1$, the growth of $i$ can be approximated as
\begin{align}
g_i(Y_i(t),t) &= \frac{\log(\sum_j A_{ij})}{t} + \frac{\log(\langle X_i(t) \rangle)}{t}.
\label{eq:discrete-growth}
\end{align}

Two implications arise from equation~(\ref{eq:discrete-growth}). First, in the transient regime there is an additive term in the growth rate which is solely dependent on the network structure. Hence, during this regime, individuals which are better connected in terms of $\sum_j A_{ij}$ should have faster growth rates. The second observation is that the second term on the right-hand side (RHS) of equation~(\ref{eq:discrete-growth}) eventually converges to the same value for each individual. This is because we study a \textit{connected} graph where participation in a pool implies that there is a path between any pair of individuals. Due to this interconnectedness, we expect that the long run time-average growth of each $Y_i$ will be dominated by the growth of the wealthiest individual in the network. 

The convergence of the long run growth rates between individuals provides a direct equivalence with the time-average growth rate $g(\langle Y \rangle_{\mathcal{N}},t) = \frac{ \mathrm{d} \log(\langle Y \rangle_{\mathcal{N}})}{\mathrm{d}t}$, which is derived from the partial ensemble average $\langle Y \rangle_{\mathcal{N}}$. This object is constructed from all individuals present in the network. As a consequence, one can use It\^{o}'s lemma to directly calculate the time-average growth rate in the network. Formally, the lemma states that the differential of an arbitrary one-dimensional function $f(\mathbf{Y},t)$ governed by an It\^{o} drift-diffusion process (such as equation~(\ref{eq:network-gbm})), is given by
\begin{align}
\mathrm{d}f(t,\mathbf{Y}) &= \frac{\partial f}{\partial t} \mathrm{d}t + \sum_i \frac{\partial f}{\partial Y_i}\mathrm{d}Y_i + \frac{1}{2}\sum_i \sum_j \frac{\partial^2 f}{\partial Y_i \partial Y_j} \mathrm{d}Y_i \mathrm{d}Y_j.
\label{eq:ito-lemma}
\end{align}
In this case, $f(t,\mathbf{Y}) = \log(\langle Y \rangle)$. The first and second derivative of $f$ in terms of $Y_i$ and $Y_j$ are
%\begin{align}
$\frac{\partial f}{\partial Y_i}  = \frac{1}{N} \frac{1}{\langle Y \rangle}$
%\label{eq:derivative-ensemble}
%\end{align}
and
%\begin{align}
$\frac{\partial^2 f}{\partial Y_i \partial Y_j} = - \frac{1}{N^2} \frac{1}{\langle Y \rangle^2}.$
%\label{eq:scnd-derivative-ensemble}
%\end{align}

Moreover, this transformation makes the differential $\mathrm{d}f(\mathbf{Y},t)$ ergodic, and since we are looking at long-run averages, $\mathrm{d}Y_i$ and $\mathrm{d}Y_i \mathrm{d}Y_j$ can be substituted with their expected values $\langle \mathrm{d}Y_i \rangle$ and $\langle \mathrm{d}Y_i \mathrm{d}Y_j \rangle$. To estimate these expectations we utilize the independent Wiener increment property $\langle \mathrm{d}W_i^2 \rangle = \mathrm{d}t$, and make use of the fact that $\sum_k A_{kj} = 1$. Further, we omit terms of order $\mathrm{d}t^2$ as they are negligible. As a result, we obtain that $\langle \mathrm{d}Y_i \rangle = \left[(1+\mu) \sum_j A_{ij} Y_j - Y_i\right] \mathrm{d}t$
and 
$\langle \mathrm{d}Y_i \mathrm{d}Y_j \rangle = \sigma^2 \mathrm{d}t \sum_k A_{ik} A_{jk} Y_k^2$. By inserting the estimates in equation~(\ref{eq:ito-lemma}) we can approximate the time-average growth rate as
\begin{align}
g_i(Y_i(t),t) &= \mu - \frac{\sigma^2}{2} \frac{\langle \gamma^2 \rangle}{N},
\label{eq:network-growth-rate}
\end{align}

where $\gamma_i = Y_i / \langle y \rangle$ is the rescaled accumulated payoff of individual $i$. This is a dimensionless quantity which compares the endowment of resources of an individual with the population average and as such has been particularly useful in analyses related to wealth inequality~\cite{bouchaud2000wealth}. In fact, equation~(\ref{eq:network-growth-rate}) indicates that the variance of the rescaled accumulated payoffs $\langle \gamma^2 \rangle_{\mathcal{N}}$ dictates the time-average growth rate. Under this model, networks with larger payoff inequality, i.e. higher $\langle \gamma^2 \rangle_{\mathcal{N}}$, are expected to have lower long-run growth rates than those where the payoffs are distributed more equally. This finding is in accordance with economics studies which suggest that wealth inequality decreases the development of an economy~\cite{herzer2012inequality,berg2011equality}.

\subsubsection{Equilibrium properties}

When deriving the individual growth rate we utilized an equilibrium property of the system. Such properties
are key to understanding the role of complex networks within the pooling and sharing mechanism. In particular, notice that in the limit we can substitute the product of $Y_j(t)$ and the exponential of~(\ref{eq:network-growth-rate}) for each $\bar{Y}_j(t+ \Delta t)$, divide both sides of the equation by the population average accumulated payoff and conclude that the equilibrium rescaled accumulated payoff of individual $i$ is
\begin{align}
\lim_{t \to \infty} \gamma_i (t) &= v_i,
\label{eq:vi}
\end{align}
where $v_i$ is the $i$-th element of the left-eigenvector of $\mathbf{A}$ associated with the largest eigenvalue normalized in a way such that $\sum_i v_i = N$.
A direct corollary is the equilibrium individual growth rate 
\begin{align}
\lim_{t \to \infty} g_i(Y_i(t),t) = \mu - \frac{\sigma^2}{2} \frac{\langle v^2 \rangle}{N}.
\label{eq:individual-growth-rate}
\end{align}
We emphasize that the quantity on the RHS of equation~(\ref{eq:individual-growth-rate}) is always greater than the time-average growth rate in equation~(\ref{eq:time-average-growth}). This can be concluded by examining the optimization problem of maximizing $\langle v^2 \rangle$ constrained on $\sum_i v_i = N$,  and noting that the global maximum is always less than $N$. Therefore a network of pooling and sharing individuals on the long run will always outperform non-cooperating GBM trajectories. 
While this indicates that cooperation is a dominant trait in the population, it also asserts that, depending on the distribution of $v$, pooling and sharing may produce societies where the distribution of accumulated payoffs differs to a great extent from the one observed in individual trajectories.

\subsubsection{Experiments}

To evaluate the differences between the distinct types of complex networks we conduct two experiments on four types of random graphs Random regular graph (RR)~\cite{bollobas2013modern}, Erdos-Renyi Poisson graph (ER)~\cite{erdos1960evolution}, Watts-Strogatz small world network (WS)~\cite{watts1998collective} and the Barabasi-Albert scale-free network (BA)~\cite{barabasi1999emergence}. Each of the graphs, except the WS graph, was described in Chapter~\ref{sec:one}. The WS graph lies between the ER and RR graphs as for its construction first one generates an RR graph, and afterwards each link is rewired with a constant probability. Moreover, each of these graphs is unipartite, which implies that every individual also represents a pool through which the accumulated payoffs are shared. 

\begin{figure}[t!]
\includegraphics[width=8.6cm]{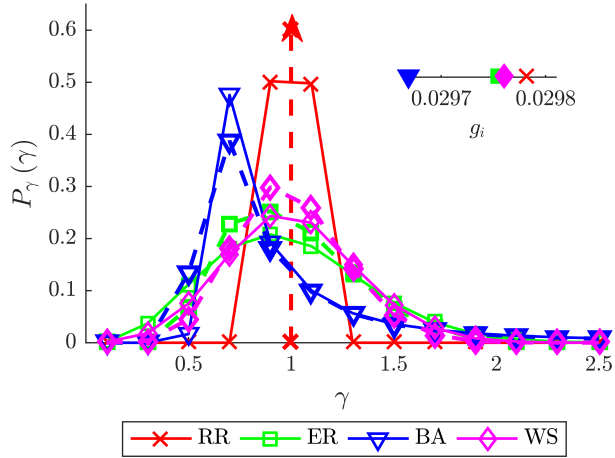}
\caption{\textbf{PDF for the rescaled accumulated payoffs.} Estimated PDF for the rescaled accumulated payoffs for four different types of random graphs -- RR, ER, WS and BA, each having an average degree $\langle d \rangle = 5$. Filled lines represent the simulated values while the dashed lines are the analytical solutions of the corresponding variables. In the simulation $\mu = 0.03$ and $\sigma^2 = 0.04$. For each graph type, the results are averaged across $100$ realizations. The graphs contain $100$ individuals involved in pooling and sharing. \label{fig:histogram}}
\end{figure}

In the first experiment, we compare the  distribution of the rescaled accumulated payoffs in equilibrium, $P_\mathrm{\gamma} (\gamma)$, among the graphs. Samples of the corresponding probability density functions (PDFs) are depicted in Fig.~\ref{fig:histogram}. We notice the agreement between the analytical solution in~(\ref{eq:vi}) (the value of $v_i$) and the simulated rescaled accumulated payoffs, $\hat{y}_i$. We observe that the RR graph exhibits no inequality across the accumulated payoffs (point mass PDF), whereas the PDFs of ER and WS graphs have exponential tails. Finally, the rescaled accumulated payoffs distribution in the BA graph resembles a fat tail. As a consequence, the BA graph has the lowest long run growth rate, followed by ER and WS, as depicted in the inset plot in Fig.~\ref{fig:histogram}.

\begin{figure}[t!]
\includegraphics[width=8.6cm]{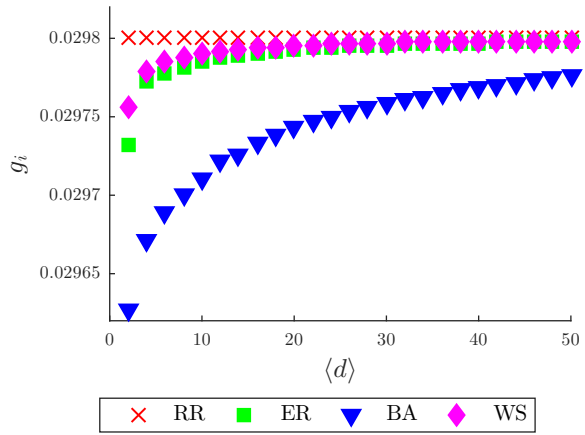}
\caption{\textbf{Network sparsity and the time-average growth rate.} The time-average growth rate as a function of the average degree $\langle d \rangle$ for the same graphs as in Fig.~\ref{fig:histogram}. Filled lines represent the simulated values while the dashed lines are the analytical solutions of the corresponding variables. In the simulation $\mu = 0.3$ and $\sigma^2 = 0.4$. For each graph type, the results are averaged across $100$ realizations. The graphs contain $100$ individuals involved in pooling and sharing. \label{fig:sparsity}}
\end{figure}

The second experiment investigates the role of network sparsity (measured through the average degree $\langle d \rangle$), on the wealth distribution. Fig.~\ref{fig:sparsity} depicts  the variance of rescaled wealth $\langle \gamma^2 \rangle$ as function of $\langle d \rangle$. The inset plot gives the ratio of the individual growth rate and the drift parameter, as function of the same variable. We observe that denser ER, WS and BA graphs yield more equal accumulated payoffs distribution compared to their respectively sparser counterparts, whereas in the RR graph the time-average growth is invariant to the average degree. 

\subsection{Discussion}

Our findings suggest that interactions on complex networks play a critical role in the observed time-average growth rates and the accumulated payoffs distribution, both in transient regime and in equilibrium. The cooperation dynamics is dictated by the properties of the underlying bipartite graph which models the network interactions in the pooling and sharing mechanism.

A startling example is the dynamics taking place on a BA scale-free graph, where largest discrepancies between the individual growth-rates are observed in the transient regime, as compared to ER, WS and RR graphs. Furthermore, the BA graph has the smallest time-average growth once the equilibrium is reached, and the most unequal accumulated payoffs distribution. From an evolutionary perspective, a network structure which presents with lower time-average growth may be interpreted as being less supportive to cooperation. It is intriguing whether there is any relationship between the apparent lower propensity to cooperation of BA scale-free networks (under the here considered interaction model) and the recent empirical evidence regarding the low-prevalence (i.e.~rarity) of scale-free networks in nature~\cite{broido2018scale,clauset2009power}.

Besides providing a basic model of self-reproducing living entities with temporal fluctuations, multiplicative processes are also excessively used to model self-financing
investments~\cite{peters2011optimal}, gambles~\cite{peters2016evaluating} and wealth allocation~\cite{berman2017empirical,bouchaud2000wealth}. In this respect, our findings may provide insights to economic utility theory with applications to finance, portfolio management, risk-evaluation and decision-making. In addition, they contribute to the ongoing discussions in economics and econophysics regarding the potential negative effects of wealth inequality on economic growth and development~\cite{bouchaud2000wealth,herzer2012inequality} and on the individual well-being in general.

As a takeaway, we conclude that inequality may arise as a result of the interwoven relationship between complex networks and cooperative dynamics in fluctuating environments. While it is known that certain network topologies promote inequality~\cite{barabasi1999emergence,salganik2006experimental}, the effect of cooperative behavior in structured populations is still to be determined~\cite{nishi2015inequality,chiang2015good,tsvetkova2018emergence}. As such, our investigations aim at providing deeper understanding on the nature of the relationship between these two occurrences.

\subsection{Generalized reciprocity in fluctuating environments}

So far, we investigated the simplest case in which every individual has the same drift $\mu$ and noise amplitude $\sigma$. In the general case, the individuals may exhibit different capabilities and have disparate luck, corresponding to heterogeneous drifts and amplitudes~\cite{peters2015evolutionary}. These situations lead to emergence of thresholds after which cooperation is no longer a dominant trait, and even partial cooperation might be observed. In this aspect, our generalized reciprocity update rule may help to ease the promotion of cooperation. However, because of the fluctuating environment, one cannot use the exact rule. A slight modification is required in order to take into account for the non-ergodic dynamics of the accumulated payoffs.

The generalized reciprocity rule in a heterogeneous environment reads
\begin{align}
    p_i\left(t+ d t\right) &= f_{i,t}\left[g_i(Y_i(t),t)\right],
    \label{eq:update-rule}
\end{align}
where $f_{i,t}: \mathbb{R} \to \left[ 0,1 \right]$. As a special case, we use the logistic function,
\begin{align}
    f_{i,t}(\omega) &= \left[ 1 + \exp{(-\kappa_i(t)(\omega - \omega_i))} \right]^{-1},
    \label{eq:logistic-function}
\end{align}
where the midpoint $\omega_i$ is given by the time-average growth rate of $g_i(Y_i(t),t)$ without pooling and sharing. The slope of the logistic function $\kappa_i(t)$  is an unbounded and increasing function of $t$, and allows for capturing the dependence of the noise amplitude on time.

It can be analytically shown that in a fluctuating environment, in the worst case (when everyone defect unconditionally), the growth rate is equal to the individual growth rate $g_i(Y_i(t),t) = \mu_i -\frac{\sigma^2_i}{2}$. This implies that even in a fluctuating environment, generalized reciprocity maintains the property of preventing exploitation~\cite{stojkoski2019evolution}.
\newpage
\section{Conclusion}

The emergence of cooperation in complex networks precludes the existence of a specific behavioral mechanism and a particular network interaction structure~\cite{Nowak-2006five}. Against this background, in this doctoral thesis we developed a unifying framework for investigating the role of generalized reciprocity in the cooperation dynamics on complex networks. Such mechanisms have been recently discovered in natural systems which involve low-level interactions among network agents with limited processing/cognitive abilities. 

While there is significant empirical evidence for the presence of these mechanisms in a plethora of real-life systems, the role of the network structure on the cooperation dynamics under general interaction models has only recently start to shape the research in various fields. In this context, we believe that the here introduced framework provides a systematic way to study the various aspects of cooperation in complex networks. In particular, the generality of the addressed framework allows for incorporation of a wide range of social dilemmas and interaction structures in the model. 

Besides its theoretical value, we believe that the framework may also be used to quantify the cooperation dynamics in real-life systems governed by similar behavioral mechanisms. We refer to a recent experimental study which suggests that the pay-it-forward principle of generalized reciprocity is a better promoter of long-term cooperation among humans than indirect reciprocity simply because it is cognitively less demanding~\cite{baker2014paying}. In this context, our (and similar) models may provide the theoretical background for the observed long-term behavior. Indeed, the above observation may be addressed from the dynamical systems perspective, under which the application of a behavioral mechanism based on generalized reciprocity yields a unique attractor where the overall level of cooperation is maximized, while at the same time the involved individuals are prevented from exploitation.

The implementation of our results goes beyond explaining the evolution of cooperation. In particular, the introduced rule is directly related to the concept of novelty search where individuals decide their next actions on the basis of previous experience~\cite{lehman2008exploiting}. Novelty search is omnipresent in reinforcement learning and has been utilized in developing machines that efficiently mimic human behavior. In this aspect, we believe that the results discovered here behave as a building block in constructing machines which behave according to the rule of generalized reciprocity.

\newpage
\section{Appendix I: Papers Published During My PhD}

\vspace{1cm}
\subsubsection*{Promoting cooperation by preventing exploitation: The role of network structure}

\vspace{0.2cm}
Zoran Utkovski, Viktor Stojkoski, Lasko Basnarkov, and Ljupco Kocarev

\vspace{0.2cm}
Physical Review E 96, 022315 (2017)

\vspace{0.2cm}
Abstract: A growing body of empirical evidence indicates that social and cooperative behavior can be affected by cognitive and neurological factors, suggesting the existence of state-based decision-making mechanisms that may have emerged by evolution. Motivated by these observations, we propose a simple mechanism of anonymous network interactions identified as a form of generalized reciprocity—a concept organized around the premise “help anyone if helped by someone'—and study its dynamics on random graphs. In the presence of such a mechanism, the evolution of cooperation is related to the dynamics of the levels of investments (i.e., probabilities of cooperation) of the individual nodes engaging in interactions. We demonstrate that the propensity for cooperation is determined by a network centrality measure here referred to as neighborhood importance index and discuss relevant implications to natural and artificial systems. To address the robustness of the state-based strategies to an invasion of defectors, we additionally provide an analysis which redefines the results for the case when a fraction of the nodes behave as unconditional defectors.

\vspace{1cm}
\subsubsection*{Cooperation dynamics of generalized reciprocity in state-based social dilemmas}

\vspace{0.2cm}
Viktor Stojkoski, Zoran Utkovski, Lasko Basnarkov, and Ljupco Kocarev

\vspace{0.2cm}
Physical Review E E 97, 052305 (2018)

\vspace{0.2cm}
Abstract: We introduce a framework for studying social dilemmas in networked societies where individuals follow a simple state-based behavioral mechanism based on generalized reciprocity, which is rooted in the principle “help anyone if helped by someone.” Within this general framework, which applies to a wide range of social dilemmas including, among others, public goods, donation, and snowdrift games, we study the cooperation dynamics on a variety of complex network examples. By interpreting the studied model through the lenses of nonlinear dynamical systems, we show that cooperation through generalized reciprocity always emerges as the unique attractor in which the overall level of cooperation is maximized, while simultaneously exploitation of the participating individuals is prevented. The analysis elucidates the role of the network structure, here captured by a local centrality measure which uniquely quantifies the propensity of the network structure to cooperation by dictating the degree of cooperation displayed both at the microscopic and macroscopic level. We demonstrate the applicability of the analysis on a practical example by considering an interaction structure that couples a donation process with a public goods game.

\vspace{1cm}
\subsubsection*{Multiplex Network Structure Enhances the Role of Generalized Reciprocity in Promoting Cooperation}

\vspace{0.2cm}
Viktor Stojkoski, Zoran Utkovski, Elisabeth Andr{\'e}, and Ljupco Kocarev

\vspace{0.2cm}
Proc. of the 17th International Conference on Autonomous Agents and Multiagent Systems (AAMAS 2018)

\vspace{0.2cm}
Abstract: In multi-agent systems, cooperative behavior is largely determined by the network structure which dictates the interactions among neighboring agents. These interactions often exhibit multidimensional features, either as relationships of different types or temporal dynamics, both of which may be modeled as a "multiplex" network. Against this background, here we advance the research on cooperation models inspired by generalized reciprocity, a simple pay-it-forward behavioral mechanism, by considering a multidimensional networked society. Our results reveal that a multiplex network structure can act as an enhancer of the role of generalized reciprocity in promoting cooperation by acting as a latent support, even when the parameters in some of the separate network dimensions suggest otherwise (i.e. favor defection). As a result, generalized reciprocity forces the cooperative contributions of the individual agents to concentrate in the dimension which is most favorable for the existence of cooperation.

\vspace{1cm}
\subsubsection*{The role of multiplex network structure in cooperation through generalized reciprocity}

\vspace{0.2cm}
Viktor Stojkoski, Zoran Utkovski, Elisabeth Andr{\'e}, and Ljupco Kocarev

\vspace{0.2cm}
Physica A: Statistical Mechanics and its Applications 531, 121805 (2018)

\vspace{0.2cm}
Abstract: Recent studies suggest that the emergence of cooperative behavior can be explained by generalized reciprocity, a behavioral mechanism based on the principle of “help anyone if helped by someone”. In complex systems, the cooperative dynamics is largely determined by the network structure which dictates the interactions among neighboring individuals. These interactions often exhibit multidimensional features, either as relationships of different types or temporal dynamics, both of which may be modeled as a “multiplex” network. Against this background, here we advance the research on cooperation models inspired by generalized reciprocity by considering a multidimensional networked society. Our results reveal that a multiplex network structure may enhance the role of generalized reciprocity in promoting cooperation, whereby some of the network dimensions act as a latent support for the others. As a result, generalized reciprocity forces the cooperative contributions of the individuals to concentrate in the dimension which is most favorable for the existence of cooperation.

\vspace{1cm}
\subsubsection*{Cooperation dynamics in networked geometric Brownian motion}

\vspace{0.2cm}
Viktor Stojkoski, Zoran Utkovski, Lasko Basnarkov, and Ljupco Kocarev

\vspace{0.2cm}
Physical Review E 99, 062312 (2019)

\vspace{0.2cm}
Abstract: Recent works suggest that pooling and sharing may constitute a fundamental mechanism for the evolution of cooperation in well-mixed fluctuating environments. The rationale is that, by reducing the amplitude of fluctuations, pooling and sharing increases the steady-state growth rate at which individuals self-reproduce. However, in reality interactions are seldom realized in a well-mixed structure, and the underlying topology is in general described by a complex network. Motivated by this observation, we investigate the role of the network structure on the cooperative dynamics in fluctuating environments, by developing a model for networked pooling and sharing of accumulated payoffs undergoing a geometric Brownian motion. The study reveals that, while in general cooperation increases the individual steady state growth rates (i.e., is evolutionary advantageous), the interplay with the network structure may yield large discrepancies in the observed individual resource endowments. We comment possible biological and social implications and discuss relations to econophysics.

\vspace{1cm}
\subsubsection*{Evolution of cooperation in networked heterogeneous fluctuating environments}

\vspace{0.2cm}
Viktor Stojkoski, Marko Karbevski, Zoran Utkovski, Lasko Basnarkov, and Ljupco Kocarev

\vspace{0.2cm}
arXiv preprint, arXiv:1912.09205 (2020)

\vspace{0.2cm}
Abstract: Fluctuating environments are circumstances in which random variations play an essential role in the evolutionary outcome. In such environments, cooperators may coexist with defectors even without the help of an  auxiliary mechanism. However, studies on the role of fluctuating environments in promoting cooperation have so far focused on simple settings. Under these settings the population is described as a well-mixed structure of entities displaying homogeneous physical traits which are unable to utilize additional decision making mechanisms. In this paper, we develop a systematic way for investigating structured populations consisting of entities with heterogeneous characteristics. By interpreting the structure as a complex network, we perform a detailed analysis on the role of interaction structures in the evolutionary stability of cooperation in fluctuating environments. We find that in such networked fluctuating environments, the dynamics induces creation of components characterized with distinct evolutionary properties. We utilize this analysis to examine the applicability of a simple decision making rule in a variety of settings. We thereby show that the introduced rule leads to steady state cooperative behavior that is always greater than or equal to the one predicted by evolutionary stability analysis. As a consequence, the implementation of our results may go beyond explaining the evolution of cooperation. In particular, they can be directly applied in domains that deal with the development of artificial systems able to adequately mimic reality, such as reinforcement learning.

\vspace{1cm}
\subsubsection*{Economic complexity unfolded: Interpretable model for the productive structure of economies}

\vspace{0.2cm}
Zoran Utkovski, Melanie F. Pradier, Viktor Stojkoski, Fernando Perez-Cruz, Ljupco Kocarev

\vspace{0.2cm}
PloS one 13-8 (2018)

\vspace{0.2cm}
Abstract: Economic complexity reflects the amount of knowledge that is embedded in the productive structure of an economy. It resides on the premise of hidden capabilities -- fundamental endowments underlying the productive structure. In general, measuring the capabilities behind economic complexity directly is difficult, and indirect measures have been suggested which exploit the fact that the presence of the capabilities is expressed in a country’s mix of products. We complement these studies by introducing a probabilistic framework which leverages Bayesian non-parametric techniques to extract the dominant features behind the comparative advantage in exported products. Based on economic evidence and trade data, we place a restricted Indian Buffet Process on the distribution of countries’ capability endowment, appealing to a culinary metaphor to model the process of capability acquisition. The approach comes with a unique level of interpretability, as it produces a concise and economically plausible description of the instantiated capabilities.

\vspace{1cm}
\subsubsection*{Sparse three-parameter restricted Indian buffet process for understanding international trade}

\vspace{0.2cm}
Melanie F. Pradier, Viktor Stojkoski, Zoran Utkovski, Ljupco Kocarev and Fernando Perez-Cruz

\vspace{0.2cm}
Proc. of 2018 IEEE International Conference on Acoustics, Speech and Signal Processing (ICASSP 2018)

\vspace{0.2cm}
Abstract: This paper presents a Bayesian nonparametric latent feature model specially suitable for exploratory analysis of high-dimensional count data. We perform a non-negative doubly sparse matrix factorization that has two main advantages: not only we are able to better approximate the row input distributions, but the inferred topics are also easier to interpret. By combining the three-parameter and restricted Indian buffet processes into a single prior, we increase the model flexibility, allowing for a full spectrum of sparse solutions in the latent space. We demonstrate the usefulness of our approach in the analysis of countries' economic structure. Compared to other approaches, empirical results show our model's ability to give easy-to-interpret information and better capture the underlying sparsity structure of data.

\vspace{1cm}
\subsubsection*{Evidence of innovation performance in the period of economic recovery in Europe}

\vspace{0.2cm}
Elena Makrevska Disoska, Dragan Tevdovski, Katerina Toshevska-Trpchevska and Viktor Stojkoski

\vspace{0.2cm}
Innovation: The European Journal of Social Science Research, 1-16 (2018)

\vspace{0.2cm}
Abstract: This paper provides empirical evidence on the innovation performance in the European countries in the years of recovery from the global economic and financial crisis by using the CDM model of simultaneous equations. The model directly links R\&D engagement and intensity to innovation outcomes measured either as process or as product innovation, and then estimates the effectiveness of the innovative efforts leading to productivity gains. The difference between the drives of innovation systems and its influence over the productivity growth is analyzed between two different institutional settings in Europe. For that purpose a company-level dataset is used from the 2012 round of the Community Innovation Survey. The results indicate that the recent financial crisis had negative influence on the companies' willingness to innovate. The effect of the crisis led to further divergence in the innovation systems of these two institutional settings. Identifying the characteristics of the innovation systems and drivers of innovation during the turmoil shows that policy instruments on EU level should be oriented towards creation of competitive business environment, encouragement to adopt the best management techniques and organizational structures and improvement of well-functioning capital, product and labor markets.

\vspace{1cm}
\subsubsection*{Correlation patterns in foreign exchange markets
}

\vspace{0.2cm}
Lasko Basnarkov, Viktor Stojkoski, Zoran Utkovski and Ljupco Kocarev

\vspace{0.2cm}
Physica A: Statistical Mechanics and its Applications, 525, 1026-1037 (2019)

\vspace{0.2cm}
Abstract: The value of an asset in a financial market is given in terms of another asset known as numeraire. The dynamics of the value is non-stationary and hence, to quantify the relationships between different assets, one requires convenient measures such as the means and covariances of the respective log returns. Here, we develop transformation equations for these means and covariances when one changes the numeraire. The results are verified by a thorough empirical analysis capturing the dynamics of numerous assets in a foreign exchange market. We show that the partial correlations between pairs of assets are invariant under the change of the numeraire. This observable quantifies the relationship between two assets, while the influence of the rest is removed. As such the partial correlations uncover intriguing observations which may not be easily noticed in the ordinary correlation analysis.

\vspace{1cm}
\subsubsection*{The Impact of a Crisis on the Innovation Systems in Europe: Evidence from the CIS10 Innovation Survey}

\vspace{0.2cm}
Katerina Toshevska-Trpchevska, Elena Makrevska Disoska, Dragan Tevdovski and Viktor Stojkoski

\vspace{0.2cm}
European Review 27, 4, 543-562 (2019)

\vspace{0.2cm}
Abstract: The varieties of the national innovation systems among European countries are reflected in the large differences, discrepancies and sometimes unexpected results in the innovation processes and their influence on labor productivity growth. The goal of this paper is to find the differences between the drivers of the innovation systems and their influence on growth of productivity between two groups of countries with different institutional settings in the period of the financial and economic crisis in Europe. The first group consists of a selection of CEE (Central and East European) countries. The second group consists of Germany, Norway, Spain and Portugal. In order to measure the role of innovation on productivity growth we use the CDM (Crepon, Duguet and Mairesse) model of simultaneous equations. The model directly links R\&D engagement and intensity to innovation outcomes measured either as process or product innovation, and then estimates the effectiveness of the innovative effort leading to productivity gains. The company-level dataset is drawn from the Community Innovation Survey (CIS10). There is one common result for the two groups, that in general the probability for a typical firm to engage in innovation increases with its size. The other factors influencing the decision process differ. A firm’s productivity increases significantly with innovation output, but only with firms operating in Western Europe. The results for firms in Central and Eastern Europe indicate that these countries’ national innovation systems are vulnerable, and in periods of crises higher level of innovation output leads to lower labor productivity. Therefore, systemic faults in the national innovation systems result in their unsustainability, especially visible in periods of crises, as was the case in 2008–2010. When it comes to Western European countries, the financial and economic crisis did not have negative effects on their innovation systems as innovation activity resulted in higher levels of labor productivity. Regarding the CEE group of countries, the crisis influenced both the innovation process and labor productivity as a whole negatively.

\vspace{1cm}
\subsubsection*{Robust determinants of companies’ capacity to innovate: a Bayesian model averaging approach}

\vspace{0.2cm}
Mijalche Santa, Viktor Stojkoski, Marko Josimovski, Igor Trpevski and Ljupco Kocarev

\vspace{0.2cm}
Technology Analysis \& Strategic Management 31, 11, 1283-1296 (2019)

\vspace{0.2cm}
Abstract: Robustness of innovation determinants is a crucial component for the company’s capacity to innovate and is increasingly central to our understanding of country (national) innovation capacity. The large number of internal and external determinants therefore raises the question of finding/perceiving the robust determinants of companies’ capacity to innovate. By using a Bayesian Model Averaging approach, the World Economic Forum’s (WEF) Competitiveness dataset of 135 countries, 10 periods, and a total of 1.239 observations, has been analysed. From 62 explanatory determinants, 27 determinants were found to be significantly and robustly correlated with companies’ capacity to innovate. Our results show that the large number of the previously suggested innovation determinants is not robust. A holistic approach that jointly considers the internal and external determinants of CCI is proposed. A central ingredient of this approach is direct public and private financial support for performing research and development.

\vspace{1cm}
\subsubsection*{Option pricing with heavy-tailed distributions of logarithmic returns}

\vspace{0.2cm}
Lasko Basnarkov, Viktor Stojkoski, Zoran Utkovski and Ljupco Kocarev

\vspace{0.2cm}
International Journal of Theoretical and Applied Finance 22, 7, 1950041 (2019)

\vspace{0.2cm}
Abstract: A growing body of literature suggests that heavy tailed distributions represent an adequate model for the observations of log returns of stocks. Motivated by these findings, here, we develop a discrete time framework for pricing of European options. Probability density functions of log returns for different periods are conveniently taken to be convolutions of the Student’s t-distribution with three degrees of freedom. The supports of these distributions are truncated in order to obtain finite values for the options. Within this framework, options with different strikes and maturities for one stock rely on a single parameter — the standard deviation of the Student’s t-distribution for unit period. We provide a study which shows that the distribution support width has weak influence on the option prices for certain range of values of the width. It is furthermore shown that such family of truncated distributions approximately satisfies the no-arbitrage principle and the put-call parity. The relevance of the pricing procedure is empirically verified by obtaining remarkably good match of the numerically computed values by our scheme to real market data.

\vspace{1cm}
\subsubsection*{Lead–lag relationships in foreign exchange markets}

\vspace{0.2cm}
Lasko Basnarkov, Viktor Stojkoski, Zoran Utkovski and Ljupco Kocarev

\vspace{0.2cm}
Physica A: Statistical Mechanics and its Applications, 539, 122986 (2020)

\vspace{0.2cm}
Abstract: Lead–lag relationships among assets represent a useful tool for analyzing high frequency financial data. However, research on these relationships predominantly focuses on correlation analyses for the dynamics of stock prices, spots and futures on market indexes, whereas foreign exchange data have been less explored. To provide a valuable insight on the nature of the lead–lag relationships in foreign exchange markets here we perform a detailed study for the one-minute log returns on exchange rates through three different approaches: (i) lagged correlations, (ii) lagged partial correlations and (iii) Granger causality. In all studies, we find that even though for most pairs of exchange rates lagged effects are absent, there are many pairs which pass statistical significance tests. Out of the statistically significant relationships, we construct directed networks and investigate the influence of individual exchange rates through the PageRank algorithm. The algorithm, in general, ranks stock market indexes quoted in their respective currencies, as most influential. In contrast to the claims of the efficient market hypothesis, these findings suggest that all market information does not spread instantaneously.

\newpage
\section{References}
\renewcommand\refname{\vskip -1cm}
%\bibliographystyle{unsrt}
%\bibliography{coop-bib}

\end{document}